\documentclass[useAMS,usenatbib,usegraphicx]{mn2e}

\usepackage{aas_macros}

\usepackage{amsmath}
\usepackage{amssymb}

\usepackage{graphicx}

\title[Rotating VPOS and predicted proper motions]{The rotationally stabilized VPOS and predicted proper motions of the Milky Way satellite galaxies}
\author[Pawlowski \& Kroupa]{Marcel S. Pawlowski\thanks{E-mail:
mpawlow@astro.uni-bonn.de}, Pavel Kroupa\\
Argelander Institute for Astronomy, University of Bonn, Auf dem H\"{u}gel 71, D-53121 Bonn, Germany\\
}

\begin{document}
\date{Accepted 2013 July 31.  Received 2013 July 29; in original form 2013 May 31}

\pagerange{\pageref{firstpage}--\pageref{lastpage}} \pubyear{2013}

\maketitle

\label{firstpage}

\begin{abstract}
The satellite galaxies of the Milky Way (MW) define a vast polar structure (VPOS), a thin plane perpendicular to the MW disc. Proper motion (PM) measurements are now available for all of the 11 brightest, `classical' satellites and allow an updated analysis of the alignment of their orbital poles with this spatial structure. The coherent orbital alignment of 7 to 9 out of 11 satellites demonstrates that the VPOS is a rotationally stabilized structure and not only a pressure-supported, flattened ellipsoid. This allows us to empirically and model independently predict the PMs of almost all satellite galaxies by assuming that the MW satellite galaxies orbit within the VPOS. 
As a test of our method, the predictions are best met by satellites whose PMs are already well constrained, as expected because more uncertain measurements tend to deviate more from the true values.
Improved and new PM measurements will further test these predictions. A strong alignment of the satellite galaxy orbital poles is not expected in dark matter based simulations of galaxy formation. Coherent orbital directions of satellite galaxies are, however, a natural consequence of tidal dwarf galaxies formed together in the debris of a galaxy collision. The orbital poles of the MW satellite galaxies therefore lend further support to tidal scenarios for the origin of the VPOS and are a very significant challenge for the standard $\Lambda$\ cold dark matter model of cosmology.
We also note that the dependence of the MW satellite speeds on Galactocentric distance appear to map an effective potential with a constant velocity $v_{\mathrm{circ}} \approx 240\,\mathrm{km\,s}^{-1}$, to about 250\,kpc. The individual satellite velocities are only mildly radial.
\end{abstract}


\section{Introduction}
\label{sect:intro}

The dwarf galaxies of the Local Group (LG; see for example \citealt{Mateo1998} and \citealt{McConnachie2012} for reviews) pose extraordinarily important constraints on our understanding of galaxy formation and evolution in a cosmological context because of the high quality of observational data.
The satellite galaxy system of the Milky Way (MW) is strongly anisotropic. The positions of the satellites can be described by a thin (20--30\,kpc root-mean-square height) plane which is oriented approximately perpendicularly to the MW and which extends to the most distant satellite galaxies Leo I and Leo II at a Galactocentric radius of about 250\,kpc. The MW satellites therefore reside within a vast polar structure (VPOS; \citealt*{Pawlowski2012a,Pawlowski2013a}). This thin, polar structure has been discovered almost four decades ago \citep{LyndenBell1976,Kunkel1976,LyndenBell1982}, and has been a persistent feature even as more and more faint MW satellite galaxies were discovered \citep*{Kroupa2005,Metz2007,Metz2009}. When independently considering either the 11 most-luminous `classical' MW satellites or the more recently discovered fainter satellites, the planar orientation determined from both samples is essentially the same even though only the first sample is commonly considered observationally complete \citep{Kroupa2010}. In addition, globular clusters (GCs) categorized as young halo (YH) GCs trace the same thin polar structure, they therefore also belong to the VPOS (\citealt*{Keller2012},\citealt{Pawlowski2012a}).

For the VPOS to be stable over time, the satellites have to orbit within it, otherwise the structure would quickly disperse. Because we reside essentially in the centre of the structure, the line-of-sight velocities of the satellites are not indicative of a common rotational sense. If, however, the tangential motion of a satellite is known via measured proper motions (PMs), then the orientation of the orbital plane and the orbital direction of the satellite can be deduced. It can be described via the satellite's orbital pole, which is the direction of the orbital angular momentum of the satellite around the MW centre. Such an analysis was performed by \citet*{Metz2008}, who at that time had available PM measurements for eight MW satellite galaxies. They determined that within the uncertainties of the orbital poles, up to six satellites co-orbit within the same orbital plane while one (Sculptor) is counter-orbiting. This indicated a rotationally stabilized VPOS, but suffered from the small number of orbital poles and their large uncertainties. This was somewhat alleviated in the meantime by the discovery that stellar and gaseous streams of disrupted MW satellite systems (both galaxies and GCs) preferentially align with the VPOS. Such an alignment is an additional indication of rotational stabilization because the streams trace the orbital plane of their progenitor objects \citep{Pawlowski2012a}.

A similar plane of satellites has been discovered with very high significance in the satellite galaxy system of M31 \citep{Ibata2013,Conn2013}. This Great Plane of Andromeda (GPoA) consists of about half of the M31 satellite galaxies and is oriented edge-on towards the MW, such that the line-of-sight velocities of the associated galaxies reveal the common orbital sense of the GPoA. In addition, there appear to be two very symmetric planar structures containing the non-satellite dwarf galaxies in the LG, which might connect the two satellite galaxy systems \citep{Pawlowski2013a}.

The alignment of the MW satellite galaxies in a common, thin plane led to the suggestion that the satellites might be second-generation objects born from the tidal debris of a past galaxy encounter \citep{LyndenBell1976,Kroupa2005,Metz2007}. The formation of such ancient tidal dwarf galaxies (TDGs) appears to be a most natural scenario for generating thin, planar distributions of co-orbiting dwarf galaxies. TDGs form within the tidal tails expelled during galaxy encounters, so they share a common orbital plane and even counter-orbiting debris within the same orbital plane appears naturally \citep*{Pawlowski2011}. A large number of TDGs can form in a single encounter \citep{Malphrus1997,Weilbacher2002,Bournaud2008}, which is further enhanced for more gas rich galaxies such as those present in the earlier Universe \citep{Kroupa2010,Bournaud2011}. TDGs can also be long lived \citep{Kroupa1997,Recchi2007}, and dwarf elliptical galaxies have been found to be similar to ancient TDGs \citep{Dabringhausen2013}. However, TDGs are not expected to contain significant amounts of dark matter \citep{Barnes1992,Wetzstein2007}. The high mass-to-light ratios $M/L$\ inferred from the velocity dispersions of the observed MW satellites might therefore be problematic for the possible tidal origin of these objects. However, the determination of $M/L$\ is based on the assumptions that the satellites are bound, virialized systems which is not necessarily the case \citep{Kroupa1997,Klessen1998,Casas2012,Smith2013}, that the measured velocity dispersions are correct, and that the underlying dynamics is Newtonian which is also sometimes questioned \citep{Angus2008,McGaugh2010,Famaey2012,McGaugh2013}. Furthermore, even observed young TDGs show signs of missing mass \citep{Bournaud2007,Gentile2007}. For a more detailed discussion of the prospects of the scenario of a tidal origin of the MW satellites, see \citet{Pawlowski2012a}.

There are currently three types of scenarios for a tidal origin of the VPOS. Initially it was suggested that a Large Magellanic Cloud (LMC) progenitor galaxy, which was disrupted on its orbit around the MW, is the parent object of several other MW satellites \citep{LyndenBell1976,Pawlowski2011} and this suggestion needs to be revisited given the wide orbit of the LMC \citep{2013ApJ...764..161K}. More recent scenarios aim at developing an understanding of the satellite galaxy planes from an LG-wide perspective. It has been suggested that a major merger has occurred to the M31 progenitor about 9 Gyr ago \citep{Hammer2010}, resulting in the formation of TDGs of which some can have reached the MW and formed the VPOS \citep{Yang2010,Fouquet2012}. This model is also able to account for the orientation and orbital directions of the M31 satellites in the GPoA \citep{Hammer2013}. An alternative galaxy interaction distributing tidal debris on LG scales is the possibility of a past encounter between the MW and M31 \citep{Sawa2005,Pawlowski2012a,Zhao2013}.

The formation of TDGs is inevitable in every cosmological model, including the currently preferred standard cosmology based on dark energy and cold dark matter ($\Lambda$CDM). Nevertheless, the MW satellite population is most-often interpreted as consisting exclusively of primordial, dark matter dominated galaxies \citep[see for example the review of][]{Battaglia2013}. This is despite a number of serious failures of the cosmological models on satellite galaxy scales (\citealt{Kroupa2010}, \citealt{Kroupa2012a}, \citealt*{Kroupa2012b}), such as the core/cusp problem \citep{Dubinski1991}, the missing satellites problem \citep{Klypin1999,Moore1999}, the missing bright satellites \citep{Bovill2011} or the too big to fail problem \citep{BoylanKolchin2011}.

The anisotropic distribution of the satellite galaxies adds to these problems because sub-halo-based satellite galaxy populations tend to be only mildly flattened \citep{Libeskind2009,Deason2011,Wang2013,Starkenburg2013}. \citet{Wang2013} report that up to 10 per cent of their simulated satellite systems are as flat as the MW satellite system, but their comparison is flawed (see \citealt{Pawlowski2013a} for a more detailed discussion). One of their measures of flattening is seriously biased towards finding a stronger flattening for a satellite system which is more radially concentrated than another, even though both systems might have the same angular distribution. Despite this bias, none of the high-resolution cosmological simulations they analyse reproduces the observed flattening of the satellite galaxies, not even when taking the obscuration by the MW disc into account. Their reported non-zero likelihood to find the observed flattening stems exclusively from the analysis of low-resolution simulations which includes unresolved satellites. Consequently, \citet{Wang2013} admit that the spatial distribution of these satellites `is uncertain and model-dependent' and `unlikely to be a very accurate estimate of the true orbit of the satellite'. This casts doubt on the validity of investigating the flattening of the spatial positions of the unresolved model satellites.

A coherent rotation of a majority of the satellite galaxies in a thin plane further worsens the inconsistency of the VPOS with the cosmological expectation because the orbital poles of dark matter sub-haloes are distributed in all possible directions nearly isotropically \citep{Pawlowski2012b}. 
It has been suggested that a large fraction of the brightest MW satellites has been accreted as a common group and that they therefore share a common orbital direction \citep{LiHelmi2008,DOnghiaLake2008,Deason2011}. This  is inconsistent with the much larger spatial extent of observed dwarf galaxy associations compared to the VPOS \citep{Metz2009b}, with the wide spread in the estimated infall times of the satellite galaxies \citep{Rocha2012} and with the results of high-resolution simulations in which most massive satellites are accreted individually \citep{Wang2013}. Furthermore, \citet{Angus2011} find the satellite galaxy proper motions to be inconsistent with a recent infall. Similar arguments can be made for the suggestion that the accretion of dwarf galaxies along cosmological dark matter filaments is responsible for thin, co-orbiting satellite galaxy planes \citep{Libeskind2005,Zentner2005,Libeskind2011,Lovell2011}. These filaments are also found to be too wide \citep[comparable to the virial radius of the main halo; see][]{Vera-Ciro2011} and do not result in thin, co-orbiting satellite galaxy planes but only in a minor over-abundance of sub-haloes with certain orbital directions \citep{Pawlowski2012b}.

Unfortunately, the velocity information provided by cosmological simulations is still frequently ignored when comparing the flattening of simulated satellite galaxy distributions with the observed one \citep{Wang2013,Starkenburg2013}. Notable exceptions are \citet{Libeskind2009} and \citet{Deason2011}, who both report that a strong concentration of the orbital poles of the brightest satellite galaxies is not expected in $\Lambda$CDM simulations.

Updated and new PM measurements for the 11 classical MW satellites now allow us to investigate the alignment of their orbital poles with unprecedented accuracy. This reveals that the majority of the classical satellites indeed co-orbit close to a common, polar plane. In Sect. \ref{sect:datasets} we compile the currently available PM measurements for the MW satellite galaxies. These are then used in Sect. \ref{sect:orbitalpoles} to determine the orbital poles and their uncertainties for all 11 classical MW satellite galaxies, which are listed together with the position and velocity vectors. The resulting distribution is discussed and compared to modelled distributions extracted from cosmological simulations in Sect. \ref{sect:discussion}. In Sect. \ref{sect:propmopredict} the PMs of all known MW satellite galaxies are predicted by requiring them to orbit within the VPOS. The predicted ranges of PMs are provided in a table and in graphical form. We end with concluding remarks in Sect. \ref{sect:conclusion}.

\section{Proper Motions}
\label{sect:datasets}

\begin{table*}
\begin{minipage}{180mm}
 \caption{MW satellite data}
 \label{tab:satellitedata}
 \begin{center}
 \begin{tabular}{@{}lcccccccc}
  \hline
  Name & $l$ & $b$ & $r_{\sun}$ & $v_{\mathrm{los}}$ & $\mu_{\alpha} \cos \delta$ & $\mu_{\delta}$ & Type & Ref. \\
   & $[^\circ]$ & $[^\circ]$ & [kpc] & $[\mathrm{km\,s}^{-1}]$ & $[\mathrm{mas\,yr}^{-1}]$ & $[\mathrm{mas\,yr}^{-1}]$ &  &  \\
   \hline

Sagittarius &  5.6 & -14.2 & 26.3 & $140.0 \pm 2.0$ & $-2.650 \pm 0.080$ &  $-0.880 \pm 0.080$ & Ground & \citet{1997AJ....113..634I} \\
 & & & & &  $-2.830 \pm 0.200$ &  $-1.330 \pm 0.200$ & Ground & \citet{2005ApJ...618L..25D} \\
 & & & & &  $-2.750 \pm 0.200$ &  $-1.650 \pm 0.220$ & HST & \citet{2010AJ....139..839P} \\
 & & & & &  $\mathit{-2.684 \pm 0.070}$ &  $\mathit{-1.015 \pm 0.070}$ &  \it Average & \\
LMC &  280.5 & -32.9 & 50.6 & $262.2 \pm 3.4$ & $+1.910 \pm 0.020$ &  $+0.229 \pm 0.047$ & HST & \citet{2013ApJ...764..161K} \\
SMC &  302.8 & -44.3 & 64.0 & $145.6 \pm 0.6$ & $+0.772 \pm 0.063$ &  $-1.117 \pm 0.061$ & HST & \citet{2013ApJ...764..161K} \\
Draco &  86.4 & 34.7 & 75.9 & $-291.0 \pm 0.1$ & $+0.600 \pm 0.400$ &  $+1.100 \pm 0.500$ & Ground & \citet{1994IAUS..161..535S} \\
 & & & & &  $+0.329 \pm 0.064$ &  $+0.172 \pm 0.062$ & HST & Piatek et al. private comm. \\
 & & & & &  $\mathit{+0.336 \pm 0.063}$ &  $\mathit{+0.186 \pm 0.062}$ &  \it Average & \\
Ursa Minor &  105.0 & 44.8 & 75.9 & $-246.9 \pm 0.1$ & $+0.500 \pm 0.800$ &  $+1.200 \pm 0.500$ & Ground & \citet{1994IAUS..161..535S} \\
 & & & & &  $+0.056 \pm 0.078$ &  $+0.074 \pm 0.099$ & Ground & \citet{1997ASPC..127..103S} \\
 & & & & &  $-0.500 \pm 0.170$ &  $+0.220 \pm 0.160$ & HST & \citet{2005AJ....130...95P} \\
 & & & & &  $\mathit{-0.036 \pm 0.071}$ &  $\mathit{+0.144 \pm 0.083}$ &  \it Average & \\
Sculptor &  287.5 & -83.2 & 85.9 & $111.4 \pm 0.1$ & $+0.720 \pm 0.220$ &  $-0.060 \pm 0.250$ & Ground & \citet{1995AJ....110.2747S} \\
 & & & & &  $+0.090 \pm 0.130$ &  $+0.020 \pm 0.130$ & HST & \citet{2006AJ....131.1445P} \\
 & & & & &  $-0.400 \pm 0.290$ &  $-0.690 \pm 0.470$ & SRG & \citet{2008ApJ...688L..75W} \\
 & & & & &  $\mathit{+0.168 \pm 0.104}$ &  $\mathit{-0.036 \pm 0.112}$ &  \it Average & \\
Sextans &  243.5 & 42.3 & 85.9 & $224.2 \pm 0.1$ & $-0.260 \pm 0.410$ &  $+0.100 \pm 0.440$ & SRG & \citet{2008ApJ...688L..75W} \\
Carina &  260.1 & -22.2 & 105.2 & $222.9 \pm 0.1$ & $+0.220 \pm 0.090$ &  $+0.150 \pm 0.090$ & HST & \citet{2003AJ....126.2346P} \\
 & & & & &  $+0.250 \pm 0.360$ &  $+0.160 \pm 0.430$ & SRG & \citet{2008ApJ...688L..75W} \\
 & & & & &  $\mathit{+0.222 \pm 0.087}$ &  $\mathit{+0.150 \pm 0.088}$ &  \it Average & \\
Fornax &  237.1 & -65.7 & 147.2 & $55.3 \pm 0.1$ & $+0.590 \pm 0.160$ &  $-0.150 \pm 0.160$ & HST & \citet{2004AJ....128..687D} \\
 & & & & &  $+0.476 \pm 0.046$ &  $-0.360 \pm 0.041$ & HST & \citet{2007AJ....133..818P} \\
 & & & & &  $+0.480 \pm 0.150$ &  $-0.250 \pm 0.140$ & SRG & \citet{2008ApJ...688L..75W} \\
 & & & & &  $+0.620 \pm 0.160$ &  $-0.530 \pm 0.150$ & Ground & \citet{2011AJ....142...93M} \\
 & & & & &  $\mathit{+0.493 \pm 0.041}$ &  $\mathit{-0.351 \pm 0.037}$ &  \it Average & \\
Leo II &  220.2 & 67.2 & 233.3 & $78.0 \pm 0.1$ & $+0.104 \pm 0.113$ &  $-0.033 \pm 0.151$ & HST & \citet{2011ApJ...741..100L} \\
Leo I &  226.0 & 49.1 & 253.5 & $282.5 \pm 0.1$ & $-0.114 \pm 0.029$ &  $-0.126 \pm 0.029$ & HST & \citet{2013ApJ...768..139S} \\

  \hline
 \end{tabular}
 \end{center}
 \small \smallskip
 Data for the MW satellite galaxies with measured PMs. The heliocentric galaxy positions are given in Galactic longitude $l$, latitude $b$\ and heliocentric distance $r_{\sun}$. The heliocentric velocities are given as the line-of-sight velocity $v_{\mathrm{los}}$\ relative to the Sun and the two components of the PM, $\mu_{\alpha} \cos \delta$\ and $\mu_{\delta}$. Type indicates whether the PM is determined from ground-based observations (ground), from space-based \textit{HST} observations (HST), via the stellar redshift gradient (SRG) method or as an uncertainty-weighted average over the other listed values for this satellite galaxy (average). The last column gives the reference for the respective PM measurement.
\end{minipage}
\end{table*}

In the following analysis we adopt the PM measurements collected from the literature as compiled in Table \ref{tab:satellitedata}. If more than one PM is listed for a satellite galaxy in Table \ref{tab:satellitedata}, we will use the error-weighted average of these values for the determination of the orbital poles in Sect. \ref{sect:orbitalpoles}.
The list includes PMs for each of the 11 brightest ('classical') MW satellite galaxies. Compared to the analysis of \citet{Metz2008} our knowledge about the PMs of the MW satellites has improved significantly. There have been updated PM measurements for several MW satellite galaxies and in addition the first PM estimates have been published for the three remaining `classical' satellites Leo I, Leo II and Sextans.

For the LMC and Small Magellanic Cloud (SMC) we adopt the most recent PM measurements by \citet{2013ApJ...764..161K}, based on a 7 yr baseline covered by three epochs of \textit{Hubble Space Telescope} (\textit{HST}) data. The resulting PMs are very precise and the uncertainties are dominated by limitations in our understanding of the internal kinematics and geometry of the LMC and SMC, according to \citet{2013ApJ...764..161K}.

Compared to the PM list compiled by \citet{Metz2008}, we also include a new \textit{HST}-based PM measurement for Sagittarius \citep{2010AJ....139..839P} and an additional ground-based PM measurement for Fornax \citep{2011AJ....142...93M}. For Draco, \citet{2008glv..book..199P} have published a new PM estimate based on a preliminary analysis of \textit{HST} observations but explicitly warned against using this value due to its uncertain nature. Instead of ignoring this warning, we have acquired an improved preliminary PM for this satellite galaxy from Piatek \& Pryor (private communication), as listed in Table \ref{tab:satellitedata}. The first PM estimates have also been published for the two most distant MW satellites Leo I \citep{2013ApJ...768..139S} and Leo II \citep{2011ApJ...741..100L}. Both are based on \textit{HST} observations.

An additional, non-astrometric method to determine satellite galaxy PMs which does not require long temporal baselines has been employed by \citet*{2008ApJ...688L..75W}. It is based on measuring the gradient in the redshifts of stars within a dwarf galaxy. In the absence of internal rotation of the galaxy, this gradient originates from the increasing line-of-sight components of the transverse motion for increasing distances of the stars from the satellite's centre. 
The method has already been discussed almost a century ago\footnote{We note that the `Great Debate' between Harlow Shapley and Heber Curtis took place in the same year. Then, they debated the scale of the Universe and the nature of spiral nebulae as either being part of the MW or being individual, distant galaxies.} when \citet{Hertzsprung1920} used radial velocities of nebulae in the Magellanic Clouds to estimate the space velocities of the galaxies, assuming them to be non-rotating. Later, \citet*{Feast1961} have mentioned the method in the context of the rotation of the Magellanic Clouds and \citet*{Feitzinger1977} have used it to measure the PM of the LMC while accounting for the galaxy's rotation.
Using this method, \citet{2008ApJ...688L..75W} have estimated the PM of the four MW satellites Sculptor, Sextans, Carina and Fornax. They find agreement of their Fornax and Carina PMs with independent PM measurements but disagreement for Sculptor, which might be due to a rotational component in that galaxy. We nevertheless include their data because they provide the only currently available PM estimate for Sextans.

We make use of the positions, distances and line-of-sight velocities of the MW satellite galaxies as collected by \citet{McConnachie2012}.

\section{Orbital Poles}
\label{sect:orbitalpoles}

\begin{table*}
\begin{minipage}{180mm}
 \caption{MW satellite galaxy orbital poles}
 \label{tab:satellitepoles}
 \begin{center}
 \begin{tabular}{@{}lcccccccccccc}
  \hline
  Name & $x$ & $y$ & $z$ & $v_{\mathrm{x}} $ & $v_{\mathrm{y}} $  & $v_{\mathrm{z}} $ & $l_{\mathrm{pole}} $ & $b_{\mathrm{pole}} $ & $\Delta_{\mathrm{pole}} $ & $h / 10^3$ & $\theta_{\mathrm{VPOS}}^{\mathrm{class}}$ & $\theta_{\mathrm{J6}} $ \\
 & kpc & kpc & kpc & $[\mathrm{km\,s}^{-1}]$ & $[\mathrm{km\,s}^{-1}]$  & $[\mathrm{km\,s}^{-1}]$ & $[^{\circ}]$ & $[^{\circ}]$ & $[^{\circ}]$ & $[\mathrm{kpc\,km\,s^{-1}}]$ & $[^{\circ}]$ & $[^{\circ}]$ \\
  \hline

Sagittarius &  17.1 & 2.5 & -6.4 & $233 \pm  3$ & $ 23 \pm  9$ & $222 \pm  8$ & 277.5 & -2.0 & 1.7 & $-5.3 \pm 0.2$ & $ 119.1_{-1.0}^{+1.0} $ &  $100.0$\\
LMC &  -0.6 & -41.8 & -27.5 & $-41 \pm 11$ & $-225 \pm  4$ & $234 \pm  5$ & 175.4 & -5.7 & 2.0 & $16.1 \pm 0.2$ & $ 19.3_{-1.6}^{+1.6} $ &  $3.2$\\
SMC &  16.5 & -38.5 & -44.7 & $  2 \pm 17$ & $-161 \pm 15$ & $149 \pm 13$ & 191.1 & -11.2 & 5.0 & $13.5 \pm 1.1$ & $ 33.2_{-3.9}^{+4.1} $ &  $15.8$\\
Draco &  -4.3 & 62.2 & 43.2 & $-59 \pm 22$ & $ 87 \pm 13$ & $-261 \pm 19$ & 190.4 & 9.2 & 4.6 & $20.6 \pm 1.7$ & $ 39.5_{-4.2}^{+4.3} $ &  $18.0$\\
Ursa Minor &  -22.2 & 52.0 & 53.5 & $  7 \pm 28$ & $ 89 \pm 19$ & $-186 \pm 20$ & 194.6 & -8.9 & 8.4 & $15.2 \pm 2.1$ & $ 37.0_{-6.3}^{+6.9} $ &  $18.1$\\
Sculptor &  -5.2 & -9.8 & -85.3 & $-33 \pm 44$ & $188 \pm 44$ & $-99 \pm  5$ & 7.8 & -4.3 & 12.4 & $-17.6 \pm 3.8$ & $ 145.2_{-11.6}^{+11.0} $ &  $167.1$\\
Sextans &  -36.7 & -56.9 & 57.8 & $-168 \pm 160$ & $112 \pm 134$ & $116 \pm 129$ & 202.2 & -43.8 & 47.2 & $26.5 \pm 12.8$ & $ 49.8_{-47.1}^{+47.2} $ &  $46.2$\\
Carina &  -25.0 & -95.9 & -39.8 & $-74 \pm 43$ & $  8 \pm 18$ & $ 40 \pm 40$ & 132.4 & -53.6 & 30.7 & $10.4 \pm 4.3$ & $ 45.7_{-30.5}^{+30.7} $ &  $62.4$\\
Fornax &  -41.3 & -51.0 & -134.1 & $-38 \pm 26$ & $-158 \pm 26$ & $114 \pm 12$ & 160.1 & 9.1 & 7.7 & $29.5 \pm 4.1$ & $ 21.6_{-3.0}^{+4.9} $ &  $21.1$\\
Leo II &  -77.3 & -58.3 & 215.2 & $101 \pm 128$ & $237 \pm 158$ & $117 \pm 51$ & 151.8 & -10.7 & 30.5 & $77.0 \pm 31.0$ & $ 5.5_{-5.1}^{+30.5} $ &  $26.5$\\
Leo I &  -123.6 & -119.3 & 191.7 & $-167 \pm 31$ & $-35 \pm 31$ & $ 96 \pm 23$ & 256.8 & -37.0 & 20.4 & $27.6 \pm 8.8$ & $ 90.2_{-20.3}^{+20.3} $ &  $79.7$\\

  \hline
 \end{tabular}
 \end{center}
 \small \smallskip
 Positions, velocities and derived orbital pole directions of the MW satellite galaxies for which PM information is available.
 Columns 2-4 contain the position in the three Cartesian coordinates with an origin in the Galactic Centre, whereby the $x$-axis points from the Sun towards the origin at the Galactic Centre and the $z$-axis points towards the North Galactic Pole. Columns 5-7 contain the velocity components in the same coordinate system. The orbital pole direction is expressed in Galactic longitude, $l_{\mathrm{pole}}$, and Galactic latitude, $b_{\mathrm{pole}}$. The uncertainty of this direction, $\Delta_{\mathrm{pole}}$, is oriented along the direction of the great circle perpendicular to the position of the satellite galaxy only. 
The absolute values of the specific angular momentum, $h = \left| \mathbf{r} \times \mathbf{v} \right|$, is given as a negative number for those satellites which are retrograde with respect to the average orbital pole of the MW satellites (as determined from the 8 most-concentrated poles, see Sect. \ref{subsect:discussion1}) and positive if they are prograde.
 $\theta_{\mathrm{VPOS}}^{\mathrm{class}}$\ reports the distance of the orbital pole from the plane fitted to all 11 classical satellite (i.e. the inclination between the orbital plane of each individual satellite and the plane fitting the current positions of all 11). If $\theta_{\mathrm{VPOS}}^{\mathrm{class}} > 90^{\circ}$\ then the satellite galaxy is counter-orbiting (retrograde with respect to the average orbital direction of the MW satellites). $\theta_{\mathrm{J6}}$\ is the angle between each orbital pole and the average direction of the six most-concentrated orbital poles, as used by \citet{Libeskind2009}.
\end{minipage}
\end{table*}

\begin{figure*}
\centering
 \includegraphics[width=180mm]{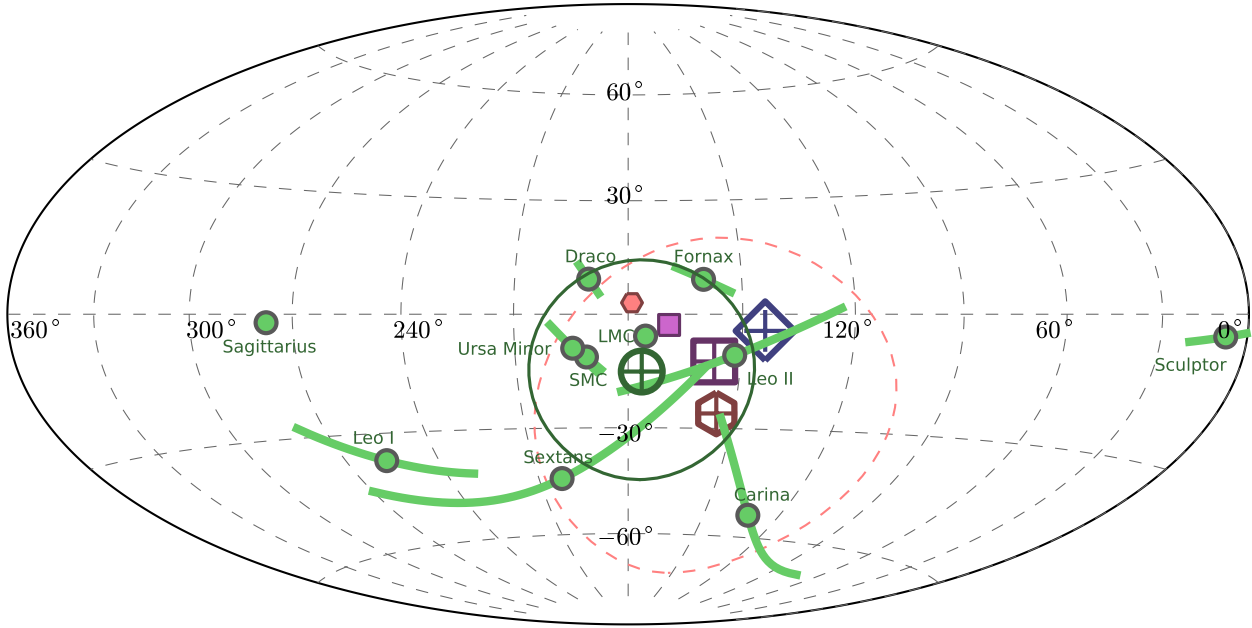}
 \caption{
 All-sky plot of the orbital poles of the MW satellite galaxies (green dots) in Galactic longitude $l$\ and Galactic latitude $b$\ in an Aitoff projection. The orbital poles are also listed in Table \ref{tab:satellitepoles} and were derived from the satellite galaxy positions, velocities and PMs compiled in Table \ref{tab:satellitedata}. The $1\,\sigma$ uncertainties of the orbital pole directions resulting from the PM uncertainties are indicated by the green great-circle segments. The average direction of the eight most concentrated orbital poles is indicated by the dark green circle with central cross at $(l,b) = (176^{\circ}.4, -15^{\circ}.0)$, the surrounding solid green circle has the size of the spherical standard distance ($\Delta_{\mathrm{sph}} = 29^{\circ}.3$) of the eight contributing poles around this direction.
The magenta square gives the direction of the normal vector to the plane fitted to the positions of the 11 brightest satellite galaxies, which all contribute an orbital pole to this plot. The blue diamond indicates the direction of the normal to the plane fitted to all YH GCs \citep{Pawlowski2012a}. The red hexagon shows the average direction of the normal vectors to all 14 streams in the MW halo analysed in \citet{Pawlowski2012a}, the dashed red circle has a size of their spherical standard distance of $\Delta_{\mathrm{sph}} = 46^{\circ}$\ and the small filled hexagon indicates the normal direction to the Magellanic Stream. The small filled square denotes the normal direction to the VPOS-3, the plane fitted to the positions of all MW satellites with the exception of three outliers \citep{Pawlowski2013a}. It will be used for the prediction of the MW satellite PMs.
The plane normal directions, the average stream normal and the majority of the orbital poles are all concentrated in the centre of the plot. Most objects in the VPOS are therefore also co-orbiting within this polar structure, Sculptor is on a retrograde orbit within the VPOS. The LMC orbital pole and the Magellanic Stream normal appear to define the centre of the distribution.
 }
 \label{fig:orbitalpolesASP}
\end{figure*}

The orbital pole, which is the direction of the angular momentum vector, is constructed for each of the 11 MW satellites using their Galactocentric positions and velocities. The orbital pole is therefore perpendicular to the orbital plane of its satellite galaxy, but in contrast to the axial nature of a normal vector, which can point in one of two opposite directions and still define the same plane, the orbital pole also illustrates the sense of rotation. Thus, if two satellite galaxies have orbital poles which are offset by $180^{\circ}$, they orbit in the same plane, but in opposite directions.

To convert the heliocentric positions and velocities of the MW satellites to the Galactocentric rest frame we adopt a Cartesian coordinate system $(x,y,z)$. The $z$-axis points towards the North Galactic Pole, the $x$-axis points in the direction from the Sun to the Galactic Centre and the $y$-axis points in the direction of the Galactic rotation. We chose the origin of the coordinate system to be the centre of the MW. We adopt $d_{\sun} = 8.3$\,kpc as the distance between the Sun and the Galactic Centre \citep{McMillan2011}.

The heliocentric PMs include both the space motion of the respective satellite galaxy and the motion of the Sun around the MW. The latter consists of the circular velocity of the local standard of rest (LSR) and the Sun's peculiar motion with respect to the LSR. For the LSR circular velocity we adopt $v_{\mathrm{LSR}} = 239\,\mathrm{km}\,\mathrm{s}^{-1}$\ \citep{McMillan2011}. For the three components of the Sun's motion with respect to the LSR we adopt the values by \citet*{Schoenrich2010}: $(U,V,W) = (11.10\,\mathrm{km}\,\mathrm{s}^{-1}, 12.24\,\mathrm{km}\,\mathrm{s}^{-1}, 7.25\,\mathrm{km}\,\mathrm{s}^{-1})$\ for the three coordinates, i.e. radially inwards to the Galactic Centre, in the direction of Galacic rotation and towards the MW north. We have checked that the resulting orbital poles do not change significantly if we adopt a LSR velocity of only $220\,\mathrm{km}\,\mathrm{s}^{-1}$, as used by \citet{Metz2008}. The orbital pole uncertainties are dominated by the PM uncertainties of the satellite galaxies \citep[see also the appendix of ][]{Metz2008}. We have therefore refrained from incorporating the uncertainties in the satellite's distance from the Sun and the uncertainties in the Sun's position and velocity relative to the MW in the following analysis.

To determine the orbital pole of a satellite galaxy, we express its position as a position vector $\mathbf{r}$\ in the Cartesian coordinate system. The resulting $(x, y, z)$\ positions are reported in Table \ref{tab:satellitepoles}. We also convert the line-of-sight velocity and PM components into the same coordinate system, resulting in a Cartesian velocity vector $\mathbf{v}$\ with components $(v_{\mathrm{x}}, v_{\mathrm{y}}, v_{\mathrm{z}})$. The orbital pole is then the direction of the angular momentum of the satellite around the MW centre.

We estimate the uncertainties of the orbital pole direction for each satellite galaxy by using a Monte Carlo technique. It randomly chooses the two PM components and the line-of-sight velocity of the satellite from Gaussian distributions centred on the most likely values of each quantity and having a width equal to the uncertainties. This is done 10\,000 times and each time the orbital pole direction and its angular distance from the most likely orbital pole are determined. We then adopt the angular distance which contains $1\,\sigma$\ (68 per cent) of the 10\,000 realizations as the $1\,\sigma$\ uncertainty of the most-likely orbital pole. This uncertainty lies on a great circle perpendicular to the direction of the satellite galaxy as seen from the Galactic Centre.

The resulting orbital pole directions, their uncertainties and the position and velocity vectors for the 11 classical satellites are compiled in Table \ref{tab:satellitepoles}. 
The table also gives the absolute value of the specific angular momentum, $h = \pm \left| \mathbf{r} \times \mathbf{v} \right|$, where $h$\ is positive if the satellite's orbital pole is prograde (i.e. the most likely orbital pole is inclined by less than $90^{\circ}$) with respect to the common pole and negative if it is retrograde. For the common pole, we have adopted the direction $(l,b) = (176^{\circ}.4, -15^{\circ}.0)$, which is the average direction of the eight most-concentrated orbital poles (see Sect. \ref{subsect:discussion1}).

Table \ref{tab:satellitepoles} also reports the inclination $\theta_{\mathrm{VPOS}}^{\mathrm{class}}$\ between the orbital pole of a satellite galaxy and the normal to the plane fitted to the positions of all 11 classical satellite galaxies. The normal of this `classical' VPOS points to $(l,b) = (157^{\circ}.1,-12^{\circ}.3)$\ and is therefore essentially the same as the best-fitting plane normals found by \citet{Metz2007} and \citet{Kroupa2010} for the same sample of galaxies. The uncertainties of $\theta_{\mathrm{VPOS}}^{\mathrm{class}}$\ indicate the maximum and minimum angles between the plane normal direction and the great circle segment representing the $1\,\sigma$\ uncertainty of the orbital pole direction. This accounts for the fact that the orbital pole direction is uncertain along a great circle only.

\section{Discussion}
\label{sect:discussion}

\subsection{A rotationally stabilized VPOS}
\label{subsect:discussion1}

\begin{figure}
\centering
 \includegraphics[width=80mm]{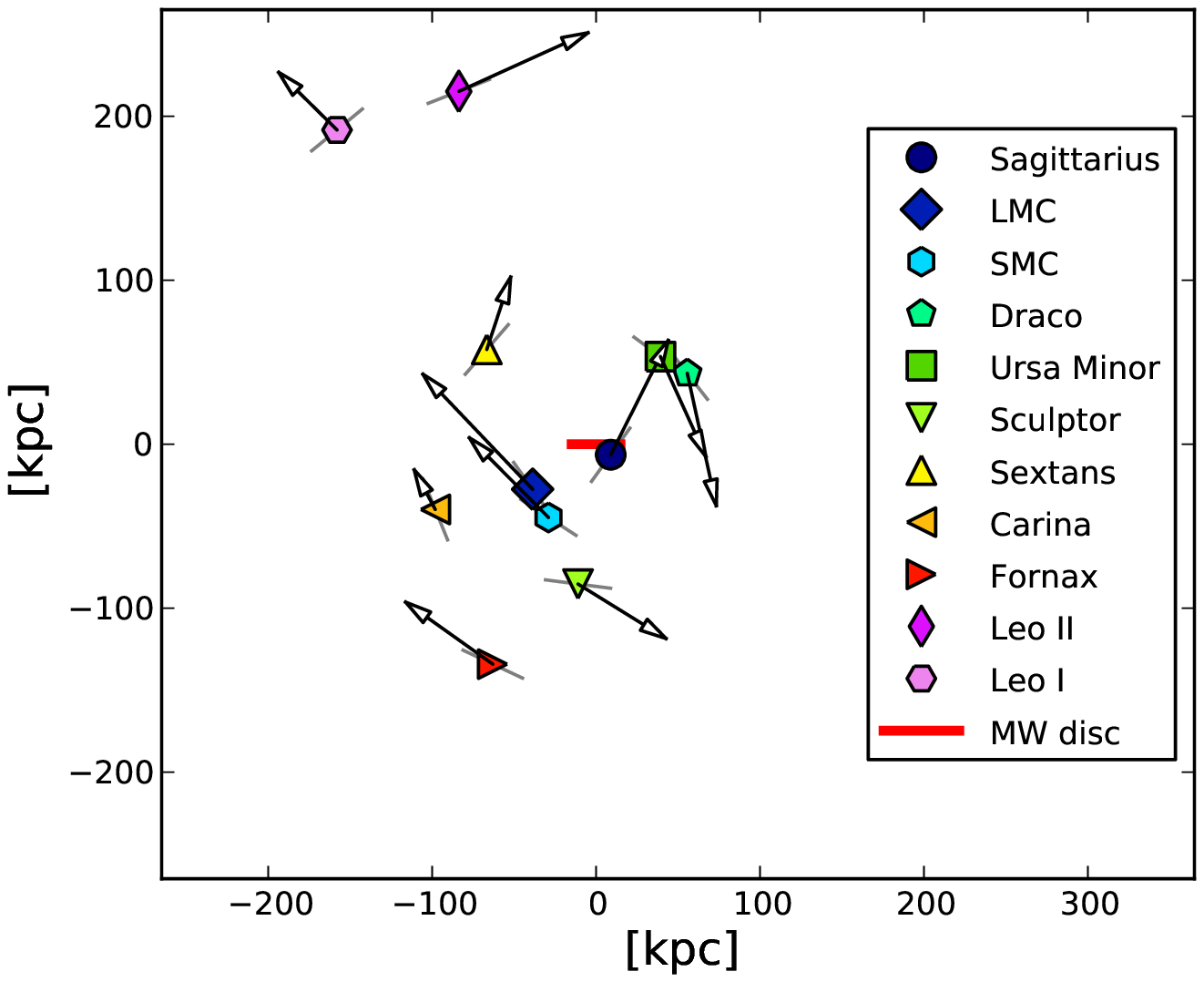}
 \caption{
 MW satellite galaxy positions and their velocities in a Cartesian projection. The MW disc is seen edge-on, as indicated by the red line. The view direction is along $l = 157^{\circ}.1$, such that the VPOS is seen approximately face-on. The velocities of the satellite galaxies in this projection are indicated by the black arrows. The length of the arrows is chosen such that 100 kpc corresponds to $25\,\mathrm{km\,s}^{-1}$. As already seen in Fig. \ref{fig:orbitalpolesASP}, the MW satellites preferentially co-orbit in a common sense (clockwise in this projection). The thin grey lines at each satellite position are perpendicular to the line connecting the MW centre and the satellite. Most velocity vectors have a small inclination to their corresponding line, so most satellite galaxies have a dominating tangential velocity component. The VPOS is therefore rotationally stabilized (compare to Fig. \ref{fig:satvel}).}
 \label{fig:faceon}
\end{figure}

\begin{figure}
\centering
 \includegraphics[width=80mm]{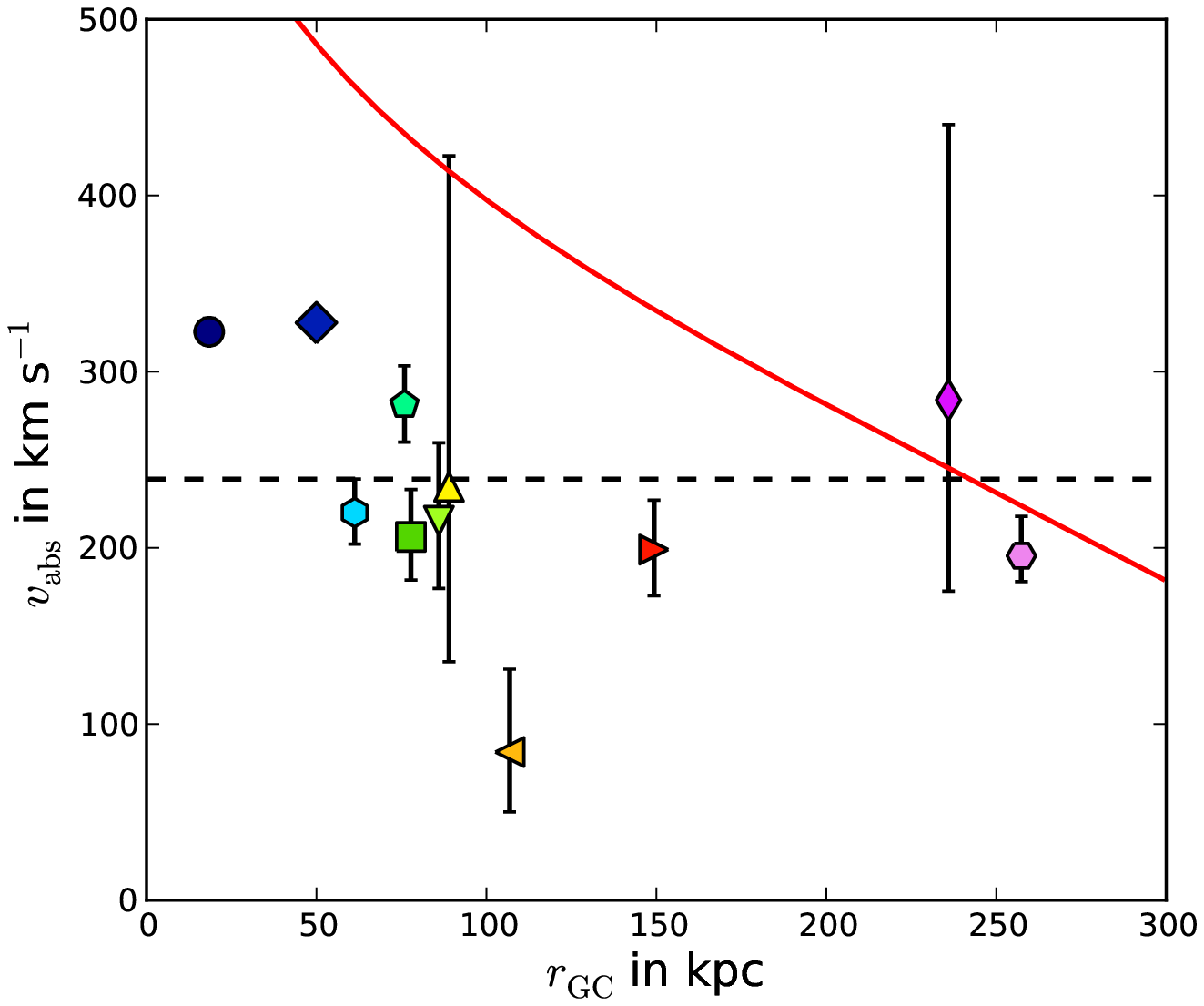}
 \includegraphics[width=80mm]{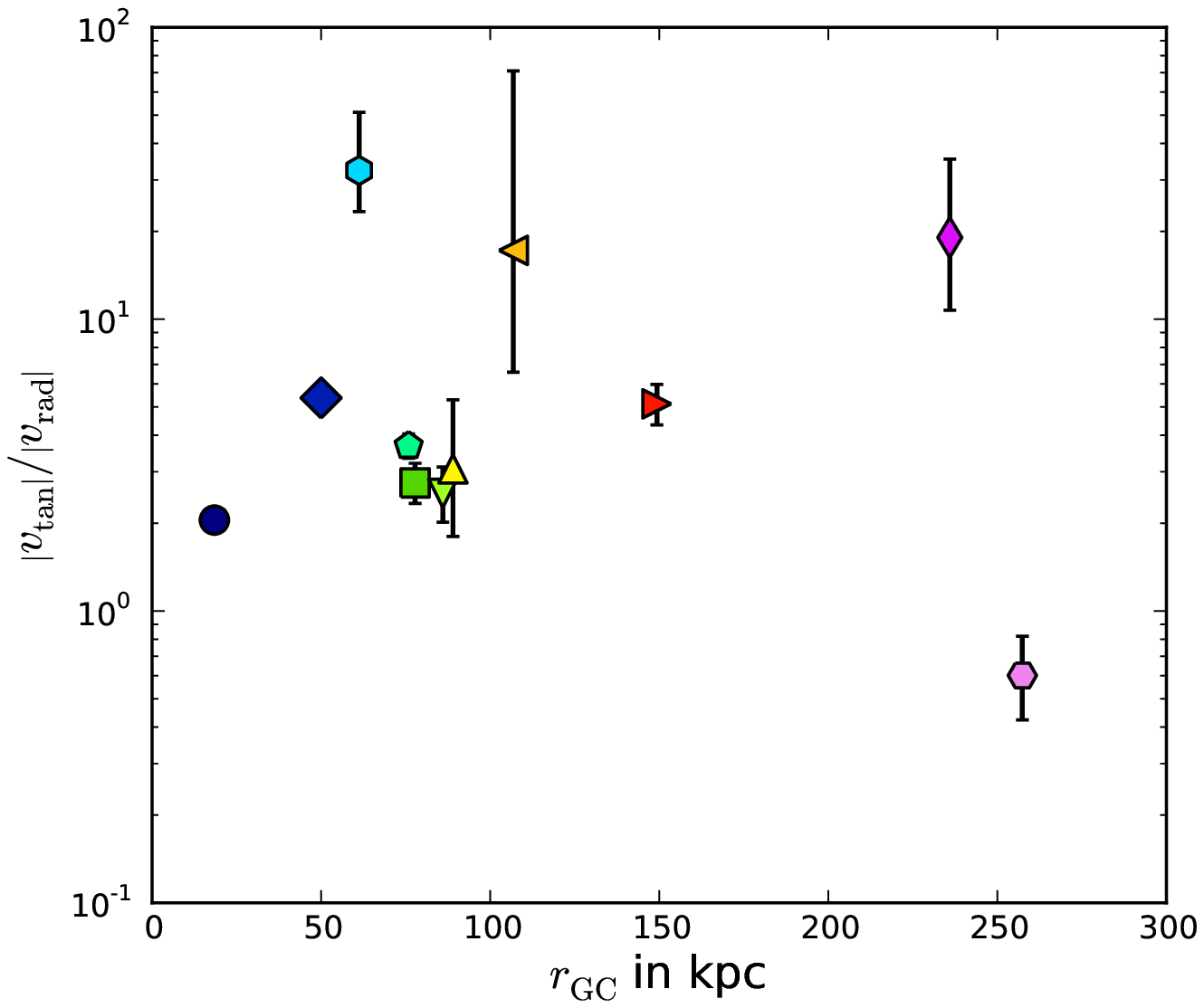}
 \caption{Velocities of the MW satellite galaxies for which PM measurements are available. Same symbols as in Fig. \ref{fig:faceon}. The \textit{upper panel} plots the absolute (three-dimensional) speeds of the MW satellite galaxies. The red line indicates the maximum velocity adopted for the PM prediction in Sect. \ref{sect:propmopredict}. If a satellite has a speed above this line it can escape from the logarithmic potential to a distance of more than 400\,kpc. The dashed horizontal line corresponds to the circular velocity of the corresponding spherical logarithmic potential ($v_{\mathrm{circ}} = v_{\mathrm{LSR}} = 239\,\mathrm{km\,s}^{-1}$), see Sect. \ref{sect:propmopredict}. Most satellites have absolute velocities which are close to this circular velocity. The innermost two satellite galaxies (Sagittarius and LMC) are close to their pericentre, where they are expected to be fastest.
The \textit{lower panel} plots the absolute value of the ratio between the tangential and the radial components of the satellite galaxy velocities relative to the Galactic Centre, $\left| v_{\mathrm{tan}} \right| / \left| v_{\mathrm{rad}} \right|$. Except for the outermost MW satellite Leo I, all satellite galaxies have a dominating tangential velocity component (compare to Fig. \ref{fig:faceon}).
 }
 \label{fig:satvel}
\end{figure}

\begin{table}
 \caption{Most concentrated orbital pole averages}
 \label{tab:concentration}
 \begin{center}
 \begin{tabular}{@{}lrrrl}
  \hline
  $N_{\mathrm{poles}}$ & $l_{\mathrm{av}} [^\circ]$ & $b_{\mathrm{av}} [^\circ]$ & $\Delta_{\mathrm{sph}} [^\circ]$ & Satellites \\
  \hline

3 &  187.0 & -8.7 & 8.5 &LMC, SMC, Ursa Minor \\
4 &  187.9 & -4.2 & 10.8 & + Draco \\
5 &  182.4 & -1.5 & 15.6 & + Fornax \\
6 &  177.3 & -3.2 & 18.5 & + Leo II \\
7 &  180.1 & -9.0 & 23.6 & + Sextans \\
8 &  176.4 & -15.0 & 29.3 & + Carina \\
9 &  182.9 & -19.2 & 35.9 & + Leo I \\
10 &  191.1 & -19.5 & 44.1 & + Sagittarius \\
11 &  191.7 & -23.1 & 62.3 & + Sculptor \\

  \hline
 \end{tabular}
 \end{center}
 \small \smallskip
The average direction of the most concentrated sample of $N_{\mathrm{poles}}$\ orbital poles out of the 11 classical satellite galaxies of the MW. It is given in Galactic longitude $l_{\mathrm{av}}$\ and Galactic latitude $b_{\mathrm{av}}$. The concentration of the orbital pole direction is measured with the spherical standard distance $\Delta_{\mathrm{sph}}$. The last column indicates which satellite galaxies contribute to each sample. For increasing $N_{\mathrm{poles}}$\ the composition of the samples changes only through the addition of more satellite galaxies.
\end{table}

The distribution of the orbital poles for the MW satellites is plotted in Fig. \ref{fig:orbitalpolesASP}, the uncertainties are plotted as green great-circle segments.  The plot also includes the direction of the normal vector to the `classical' VPOS (the minor axis of the distribution of the 11 satellites) as a magenta square and the direction of the normal vector to the best-fitting plane of the YH GCs. The red hexagon indicates the average stream normal vector, the preferred orientation of orbital planes derived from streams of stars and gas in the MW halo.

The majority of the orbital poles cluster near the centre of the plot, and also close to the MW equator which indicates polar orbits. This includes the orbital poles of the LMC and SMC, Draco, Ursa Minor, Fornax and Leo II. These satellite galaxies therefore orbit around the MW in a common direction and almost in the same orbital plane. Within their considerable uncertainties, the orbital poles of Sextans and Carina come close to this common direction, too. This preferred direction is close to the normal direction of the classical VPOS. \textit{The orbital poles of most MW satellites align with the short axis of the satellite distribution. The satellites co-orbit in the plane defined by their positions. Therefore the satellites will remain close to the VPOS plane, such that this spatial structure is rotationally stabilized and not a transient feature.} The majority of orbital poles are also close to the average stream normal, which also indicates a preferred alignment of the orbits of satellite objects in the MW halo with the planar structure consisting of satellite galaxies and YH GCs.

The orbital pole of Sculptor is offset by about $180^{\circ}$\ from the direction preferred by the other satellites, so Sculptor appears to be counter-orbiting within the same orbital plane, as has already been noticed by \citet{Metz2008}. In their search for possible streams of satellite galaxies and GCs, \citet{LyndenBell1995} point out that Sculptor lies very close to the plane of the Magellanic Stream and that it is therefore probably associated with the stream. Based on this association and several assumptions, including that all objects in a stream have the same specific angular momentum in both absolute value and direction, they predict Sculptor's PM to be $\mu_{\alpha} \cos \delta = 0.5~\mathrm{mas\,yr}^{-1}$\ and $\mu_{\delta} = - 0.8~\mathrm{mas\,yr}^{-1}$.
This PM would result in an alignment of Sculptor's orbital pole with our average orbital pole direction, but unfortunately the published PMs of Sculptor are clearly offset from this prediction. In particular, the least uncertain measurement based on \textit{HST} data by \citet{2006AJ....131.1445P}, which dominates our uncertainty-weighted average, differs from this prediction by approximately $3 \sigma$\ in $\mu_{\alpha} \cos \delta$\ and $6 \sigma$\ in $\mu_{\delta}$. A significant revision of the observed PM for Sculptor appears unlikely, but this possibility cannot be completely ruled out at the moment given the generally very uncertain nature of PM data and the wide scatter among different measurements. However, the assumption that all members of a common stream co-orbit in the same sense is not essential for a stream of tidal origin. In fact, tidal debris of interacting galaxies is naturally found to populate both co- and counter-rotating orbits \citep{Pawlowski2011}, such that a counter-orbiting object does not invalidate the underlying idea of \citet{LyndenBell1995}.

As has also already been discussed by \citet{Metz2008}, Sagittarius orbits approximately perpendicularly to the majority of the MW satellites. Its position does not allow it to orbit within the same plane (see Sect. \ref{sect:propmopredict}). It is also the satellite galaxy closest to the MW disc, such that its orbit might have suffered most from precession and possibly even strong encounters with other satellites \citep{Zhao1998}.

Leo I also appears to be an outlier, even though its orbital pole direction is still relatively uncertain. Leo I's orbital pole is also approximately perpendicular to the preferred orbital pole direction of the MW satellites and the uncertainties do not reach the preferred direction. In contrast to Sagittarius, Leo I is the most distant MW satellite galaxy, so its orbit should not have suffered from precession. Scattering due to close encounters with other satellite galaxies is also expected to be  unlikely at such a large distance. The large absolute speed of Leo I has led \citet{2013ApJ...768..139S} to argue that this galaxy might only recently have fallen in on to the MW. If this is the case, it might not be considered a satellite galaxy and could indeed be unrelated to the VPOS.
Leo I is also one of the three galaxies with the largest perpendicular distances from the VPOS fitted to all MW satellites \citep{Pawlowski2013a}. As will be discussed in Sect. \ref{fig:propmopredict}, removing Leo I and the other two outliers results in a much thinner VPOS fit which at the same time aligns better with the majority of the orbital poles.

To quantify the concentration of the orbital poles we determine the average direction $(l_{\mathrm{av}}, b_{\mathrm{av}})$\ of the $N_{\mathrm{poles}}$\ best-aligned poles on the sphere\footnote{
The inclusion of Sculptor changes the average orbital pole direction mostly along the $b$\ coordinate because we average the directions on a sphere and the average without Sculptor's pole (with $N_{\mathrm{poles}} = 10$\ in Table \ref{tab:concentration}) is essentially opposite to the direction of Sculptor's pole in $l$, but closer in $b$.
} and determine their spherical standard distance $\Delta_{\mathrm{sph}}$\ around this average direction \citep[see ][]{Metz2007}.
The results are compiled in Table \ref{tab:concentration} for combinations of 3 to all 11 poles. The eight most concentrated orbital poles have $\Delta_{\mathrm{sph}} = 29^{\circ}.3$. This has to be compared to the value derived by \cite{Metz2008}, which was $\Delta_{\mathrm{sph}} = 35^{\circ}.4$\ for only the six most-concentrated orbital poles. \textit{Thus, despite an increase in the number of orbital poles the concentration has increased by using higher-quality PM data.} If we only choose the six most-concentrated poles the concentration becomes even stronger: $\Delta_{\mathrm{sph}} = 18^{\circ}.5$.

The absolute values of the specific angular momenta of most satellite galaxies are similar, lying mostly in the range $h = 10--30~\mathrm{kpc\,km\,s}^{-1}$\ (see Table \ref{tab:satellitepoles}). In particular, most of the galaxies which \citet{LyndenBell1995} consider to be possible members of the Magellanic Stream (LMC, SMC, Draco, Ursa Minor, Sculptor and Carina) have a specific angular momentum which is close to $h = 13.0~\mathrm{kpc\,km\,s}^{-1}$, the value predicted by \citet{LyndenBell1995}. Interestingly, even Sculptor, which appears to be counter-orbiting within the VPOS, has a similar absolute value for its specific angular momentum $\left| h \right| = 17.6~\mathrm{kpc\,km\,s}^{-1}$. 
Sextans, Fornax and Leo I have specific angular momenta which are larger by about a factor of two, but their values are also less well constrained. Overall, there appears to be a trend for larger $h$\ with increasing Galactocentric distance, which is related to the similar tangential velocities of the satellites at different distances.

Fig. \ref{fig:faceon} shows the MW satellite galaxy velocities in an approximately face-on view of the VPOS. As indicated by the concentration of the orbital poles close to the VPOS normal (Fig. \ref{fig:orbitalpolesASP}), the satellites preferentially orbit in the same direction around the MW (clockwise in the figure). The plot also shows that almost all satellite galaxy velocities are dominated by motion tangential to the MW disc. The VPOS is therefore not merely a pressure-supported flattened ellipsoid, \textit{but indeed a rotationally stabilized planar structure}.

The absolute velocities of the MW satellites are plotted in the upper panel of Fig. \ref{fig:satvel}. They scatter tightly around the circular velocity of the MW disc, indicated by the dashed black line at the LSR circular velocity. The two innermost satellite galaxies (Sagittarius and the LMC) are significantly faster, but they are close to the pericentre of their orbit. Only Carina is significantly slower than the rest of the MW satellites.
The lower panel of Fig. \ref{fig:satvel} plots the ratio between the tangential and the radial velocity of the MW satellites, $\left| v_{\mathrm{tan}} \right| / \left| v_{\mathrm{rad}} \right|$. It confirms what has already been apparent in Fig. \ref{fig:faceon}: the velocities are dominated by the tangential component. Only Leo I, the outermost galaxy in the sample, is on a strongly radial orbit. All other satellites have at least a two times larger tangential than the radial velocity component.

\subsection{Comparison with model expectations}
\label{subsect:discussion2}

\begin{figure*}
\centering
 \includegraphics[width=180mm]{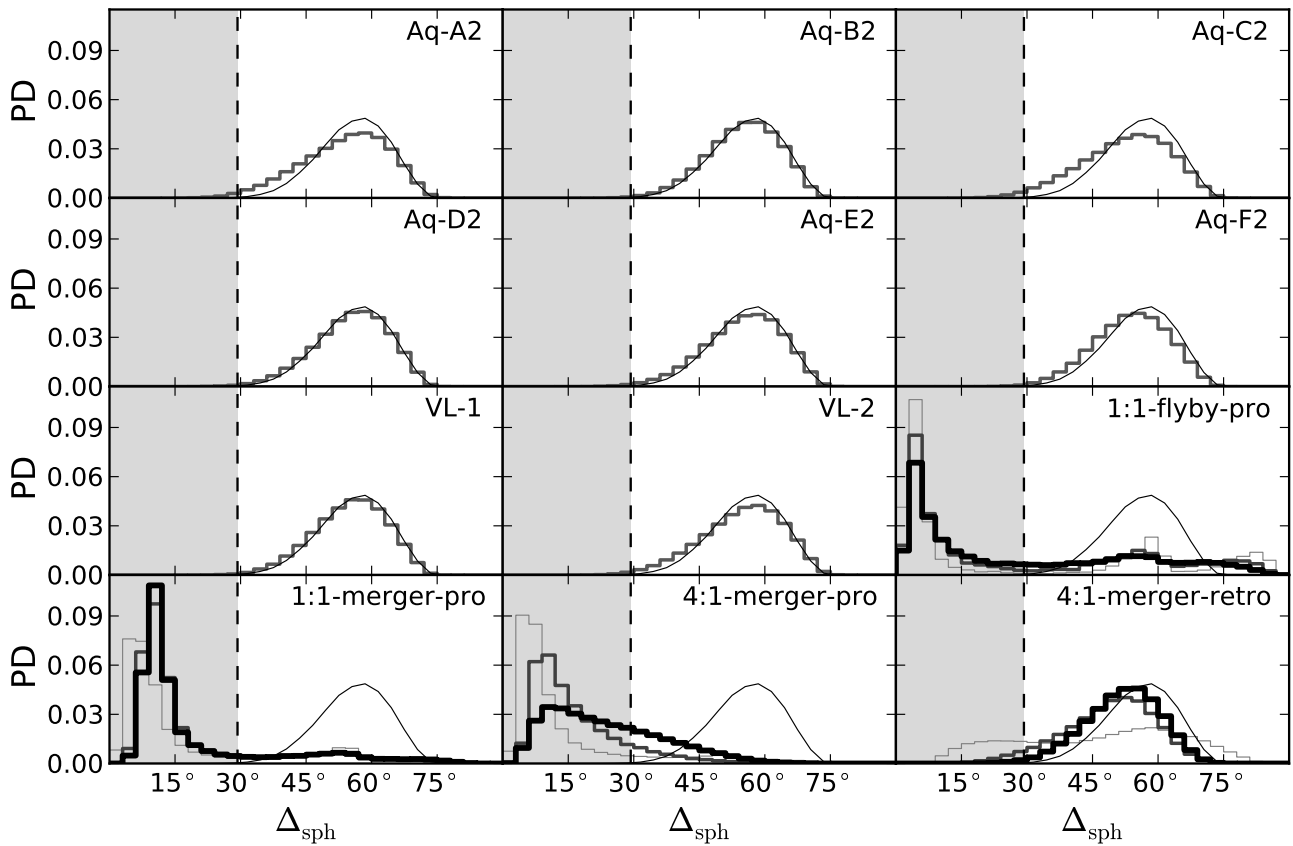}
 \caption{An updated version of fig. 3 of \citet{Pawlowski2012b}. 
 Plotted are the distributions of the probability density (PD) of the spherical standard distance $\Delta_{\mathrm{sph}}$\ of the 8 most closely aligned orbital poles out of 11 randomly drawn orbital poles. In the first eight panels, the orbital poles are drawn from the sub-halo angular momentum vectors of the cosmological high-resolution simulations Aquarius \citep{Springel2008,Lovell2011} and Via Lactea I and II \citep{Diemand2007,Diemand2008}. The last four panels show the distribution that results when drawing angular momentum vectors from the particles in calculations of galaxy interactions \citep{Pawlowski2011}. They each contain three distributions for different times after the start of the calculations of each model (thin light grey for 5\,Gyr, dark grey for 7.5\,Gyr, thick black for 10\,Gyr). The case of an isotropic sample of angular momenta is included as the thin black line in all panels for comparison. 
 The spherical standard distance $\Delta_{\mathrm{sph}} = 29^{\circ}.3$\ for the 8 best-fitting orbital poles of the MW satellite galaxies is indicated by the vertical dashed lines. To reproduce the observed clustering, a realization has to fall to the left of this line (shaded region). 
The distributions resulting from most cosmological simulations are nearly indistinguishable from the isotropic case. The vast majority of spherical standard distances determined for the randomly drawn sub-haloes in cosmological simulations are significantly larger than the observed one. This is completely different for most examples of tidal debris which have strongly concentrated distributions of angular momentum directions, in good agreement with the distribution inferred from the MW satellite galaxies.
 }
 \label{fig:DeltaSph}
\end{figure*}

To test whether the clustering of the orbital poles is a significant result we first employ the test suggested by \citet{Metz2008}. We construct the cumulative distribution of the orbital poles from the direction of the minor axis of the satellite galaxy positions. We then test this against the fitting function for the cumulative biased random pole distance distribution provided by \citet{Metz2008} using a Kuiper test \citep{Fisher1987,NumRec1992}. We can exclude the possibility that the observed cumulative pole distance distribution is drawn from the random one with 98 per cent significance, but not with 99 per cent.

This test only accounts for the overall distribution of the pole distances and depends on the direction of the minor axis of the satellite galaxy distribution. As an additional test we employ the Monte Carlo method presented in \citet{Pawlowski2012b}, their `clustering criterion'. The goal of that analysis was to test whether the orbital poles of dark-matter sub-haloes drawn from high-resolution cosmological simulations can reproduce the observed clustering of the MW satellite orbital poles. With the updated orbital poles we can now also update the results of \citet{Pawlowski2012b}.

We follow the same procedure laid out in detail in section 2.4 of \citet{Pawlowski2012b}. From a given parent distribution we randomly draw 11 orbital poles. These parent distributions are identical to those used before: either an isotropic distribution, the distribution of orbital poles of simulated sub-haloes or of tidal debris from modelled galaxy encounters. We then determine which combination of 8 of these 11 randomly selected poles results in the smallest $\Delta_{\mathrm{sph}}$\ and store this value. This is repeated 100\,000 times for each parent distribution and the resulting distribution of $\Delta_{\mathrm{sph}}$\ values is compared to the observed value of $\Delta_{\mathrm{sph}} = 29^{\circ}.3$\ for 8 out of 11 orbital poles (Fig. \ref{fig:DeltaSph})\footnote{The results do not change significantly if we test for the 6 most concentrated orbital poles out of 11.}.

An isotropic distribution of the orbital poles reproduces the observed clustering of the orbital poles of the MW satellites in only $0.10 \pm 0.01$ per cent of the cases. The observed clustering of the MW orbital poles is therefore highly significant.
Averaging over all cosmological simulations, the six Aquarius simulations \citep{Springel2008} and the two Via Lactea simulations \citep{Diemand2007,Diemand2008}, the likelihood to randomly draw a set of 11 sub-haloes of which eight are as concentrated as in the observed case is only $0.64$\ per cent. \textit{The orbital pole distributions of dark matter sub-haloes are unable to naturally reproduce the observed clustering.} The largest probability of $1.67 \pm 0.04$\ per cent is found for the Aquarius halo C2. 

\citet{Pawlowski2012b} have also compared the observed concentration of orbital poles with those of tidal debris produced in modelled galaxy interactions. They report that drawing from the modelled debris has a large probability of up to 90 per cent of resulting in similarly or more concentrated orbital pole samples. Repeating the analysis with the updated parameters (8 of 11 poles resulting in $\Delta_{\mathrm{sph}} \leq 29^{\circ}.3$\ instead of 6 of 8 resulting in $\Delta_{\mathrm{sph}} \leq 35^{\circ}.4$) leaves this result essentially unchanged. \textit{The likelihoods to reproduce the observed clustering with orbital poles drawn from tidal debris remain very high.
Compared to the sub-halo hypothesis, where the likelihood drops by about an order of magnitude when updating the analysis with the higher-quality PM data, the tidal models are unaffected.} This provides further support for the scenario in which the MW satellites have been born as TDGs in a past galaxy encounter.

The previous analysis draws orbital poles from all dark matter sub-haloes in a simulation. It might be more meaningful to compare the observed orbital pole distribution to the orbital poles of those sub-haloes which are expected to harbour the brightest satellite galaxies. Such analyses have been performed by \citet{Libeskind2009} and \citet{Deason2011}. Both studies compare the orbital poles of modelled satellite galaxies with the orientation of the minor axis of the spatial distribution of the satellites, but they both cut down their MW-like main halo sample before this comparison.

\citet{Libeskind2009} populate cosmological dark-matter-only simulations via semi-analytical galaxy formation models with luminous satellite galaxies. They select main haloes in the mass range of $2 \times 10^{11}$\ to $2 \times 10^{12}\,\mathrm{M}_{\sun}$, resulting in a sample of 30\,946 main haloes. This sample is then cut down to only 3201 isolated host haloes by rejecting all haloes containing galaxies less luminous than $M_{\mathrm{V}} = -20$, which are too faint to resemble the MW. In addition, all haloes containing less than 11 luminous satellites are rejected, resulting in a final sample of 436 host haloes whose satellite galaxy populations they then compare to the MW satellites. Therefore, only 1.4 per cent of the simulated MW-like main haloes contain a galaxy which has a satellite system that can be compared to the MW system. To determine how many main haloes in the MW halo mass range are expected to contain an MW-like galaxy (in luminosity only) that has a satellite galaxy system which is co-orbiting, the numbers of \citet{Libeskind2009} thus have to be multiplied by 0.014.

\citet{Deason2011} make use of satellite galaxies formed in hydrodynamical cosmological simulations which include baryonic physics. They require the main haloes to be in the mass range between $5 \times 10^{11}$\ and $5 \times 10^{12}\,\mathrm{M}_{\sun}$, thus including haloes which are much more massive than the one assumed around the MW. They then cut down their approximately 780 haloes in this mass range by requiring them to be relaxed and to contain a central galaxy with a disc-to-total stellar mass ratios of at least 0.3, so this again includes galaxies which are very different from the disc-dominated MW (and M31). Finally, only approximately 80 of these host haloes contain at least 10 satellites. Of the initial mass-selected sample of haloes, thus only 10 per cent remain after applying the cuts. To determine how likely it is to have similarly concentrated satellite galaxy orbital poles around a MW-like galaxy, the numbers of \citet{Deason2011} therefore need to be multiplied by at least 0.1. This can only be considered an upper limit because their final galaxy sample likely includes satellite systems around host galaxies which have properties (halo mass, disc fractions) that are very different from the MW. \citet{Deason2011} then compare the orbital poles of the 10 most-massive model satellites with the minor axis of the spatial distribution of the satellites. In their comparison with the classical satellites of the MW they neglect Sagittarius\footnote{See for example the lower panel of their fig. 4, where the innermost MW satellite is located at $\approx 50$\,kpc from the MW centre.}, so for consistency we also need to ignore this object when we compare the observed orbital poles with the \citet{Deason2011} results.

For the observed MW satellite orbital poles, 7 poles are closer than $45^{\circ}$\ from the minor axis defined by the satellite positions. According to fig. 9 of \citet{Libeskind2009}, this number is found in $\approx 5$\ per cent of their selected satellite galaxy populations, while according to fig. 8 of \citet{Deason2011} this is found in $\approx 5$\ per cent of their selected models, too. When accounting for the cuts that lead to the final satellite samples which \citet{Libeskind2009} and \citet{Deason2011} use for their figures, the likelihood to find an MW-like galaxy in an MW-like halo to be surrounded by an MW-like satellite population which has well-concentrated orbital poles aligned with the satellite system's minor axis is less than $0.05 \times 0.014 = 7 \times 10^{-4}$\ to $0.05 \times 0.1 = 5 \times 10^{-3}$. Within their $1\,\sigma$\ uncertainties, up to 9 (7) of the observed orbital poles of MW satellites can be within $45^{\circ}$\ ($30^{\circ}$) of the short axis of their spatial distribution. Such high numbers of aligned orbital poles can not be reproduced by any of the selected satellite galaxy populations of \citet{Libeskind2009} and \citet{Deason2011}.

\citet{Libeskind2009} also compare the individual orbital poles of the modelled satellite galaxies with the direction of the six most concentrated orbital poles, $J_6^{\mathrm{mean}}$. 
For the observed MW satellites, $J_6^{\mathrm{mean}}$\ corresponds to the fourth row in Table \ref{tab:concentration}. We have calculated the angle between this direction and all orbital poles (last column in Table \ref{tab:satellitepoles}). Of the 11 classical satellites, 6 are within $26^{\circ}.5$ of $J_6^{\mathrm{mean}}$. The next nearest pole is that of Sextans at an angular distance of $46.2^{\circ}$, but within its uncertainty it could be closer than $30^{\circ}$\ to $J_6^{\mathrm{mean}}$. In addition, Sculptor's orbital pole is also closely aligned with $J_6^{\mathrm{mean}}$\ ($13^{\circ}$\ inclined), but pointing in the opposite direction. According to figure 9 of \citet{Libeskind2009}, it is very unlikely ($\approx 1$\ per cent of their selected satellite systems) that 6 simulated satellites align with $J_6^{\mathrm{mean}}$\ by less than $30^\circ$. In none of their selected satellite galaxy systems do seven satellite orbital poles align better than even a larger angle of $45^{\circ}$\ with $J_6^{\mathrm{mean}}$, but this number is well possible for the observed MW satellite system.

These comparisons indicate that the observed distribution of the orbital poles of the MW satellite galaxies is extremely unlikely if they are dark matter sub-haloes as modelled in $\Lambda$CDM simulations. \textit{Within the observed orbital pole uncertainties it is by now rather certain that this hypothesis is already ruled out.} More precise PM measurements, in particular for the satellite galaxies Sextans and Carina, will help to cement this verdict.

\section{Predicting PMs}
\label{sect:propmopredict}

\begin{figure}
\centering
 \includegraphics[width=88mm]{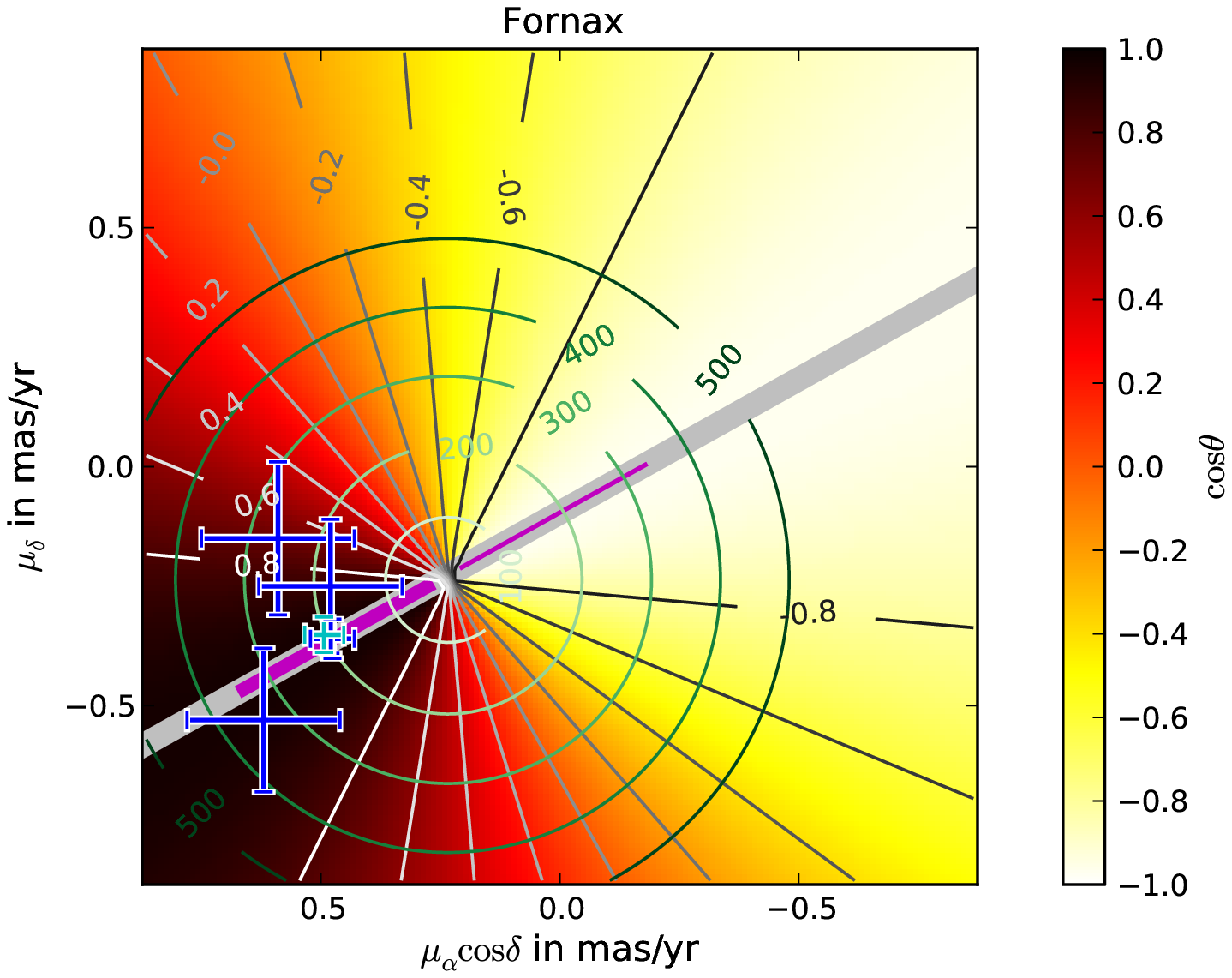}
 \caption{
Predicting the PM of the MW satellite Fornax. The two PM components are represented by the horizontal ($\mu_{\alpha} \cos \delta$) and vertical ($\mu_{\delta}$) axes. For each combination of the two PM components the colour map illustrates the angle $\theta$\ between the VPOS-3 plane and the orbital plane which would result from the PM. The radial grey contour lines also illustrate this angle, which is measured in $\cos \theta$. All PM within the $\cos \theta > 0.8$\ ($\cos \theta < -0.8$) contour result in orbital planes aligning to better than $37^{\circ}$\ with the VPOS-3 and are therefore considered to be co-orbiting (counter-orbiting). The circular green contours indicate the absolute speed of the satellite galaxy relative to the MW in steps of $100\,\mathrm{km\,s}^{-1}$. The broad grey line represents those PMs which would result in the best-possible alignment of the satellite galaxy's orbital plane and the VPOS-3. The thick magenta line indicates the predicted PM if the satellite is co-orbiting, the thin magenta line indicates the counter-orbiting case. They are truncated at the minimum and maximum speed assumed at the position of the satellite in the logarithmic halo of the MW.
The blue error bars are the individual PM measurements as compiled in Tabel \ref{tab:satellitedata}, the cyan error bar corresponds to the uncertainty-weighted average of these measured PMs. The error bars are close to the thick magenta line, therefore the measured PMs agree with the predicted (co-orbiting) one.
 }
 \label{fig:propmopredict}
\end{figure}

\begin{table*}
\begin{minipage}{180mm}
 \caption{Predicted PMs of MW satellite galaxies assuming them to orbit within the VPOS-3}
 \label{tab:prediction}
 \begin{center}
 \begin{tabular}{@{}lccccccccc}
  \hline
  Name & $l$ & $b$ & $r_{\sun}$ & $v_{\mathrm{los}}$ & $\theta_{\mathrm{VPOS-3}}^{\mathrm{predicted}}$ &  $\theta_{\mathrm{VPOS-3}}^{\mathrm{measured}}$ & $[v_{\mathrm{min}},~v_{\mathrm{max}}]$ & 
 $\begin{pmatrix} \mu_{\alpha} \cos \delta \\ \mu_{\delta} \end{pmatrix}_{\mathrm{co}}$ & 
 $\begin{pmatrix} \mu_{\alpha} \cos \delta \\ \mu_{\delta} \end{pmatrix}_{\mathrm{counter}}$ 
 \\
   & $[^\circ]$ & $[^\circ]$ & [kpc] & $[\mathrm{km\,s}^{-1}]$ & $[^\circ]$ & $[^\circ]$ & $[\mathrm{km\,s}^{-1}]$ & 
 $[\mathrm{mas\,yr}^{-1}]$ & 
 $[\mathrm{mas\,yr}^{-1}]$ \\

 \hline

Sagittarius &  5.6 & -14.2 &  26 & $ 140 $ & $ 60.2 $ & $ 108.0_{-0.7}^{+0.7} $ &  [207,~548] & no prediction & no prediction \\
Segue &  220.5 & 50.4 &  23 & $ 208 $ & $ 36.1 $ & no PM &  [126,~592] & $\begin{pmatrix} [-0.36,~+2.37] \\ [-1.43,~+2.49] \end{pmatrix}$ & $\begin{pmatrix} [-1.03,~-3.76] \\ [-2.39,~-6.31] \end{pmatrix}$\\
Ursa Major II &  152.5 & 37.4 &  32 & $ -116 $ & $ 55.4 $ & no PM &  [50,~529] & no prediction & no prediction \\
Bootes II &  353.7 & 68.9 &  42 & $ -117 $ & $ 12.8 $ & no PM &  [139,~499] & $\begin{pmatrix} [-0.48,~+1.10] \\ [-0.81,~+0.55] \end{pmatrix}$ & $\begin{pmatrix} [-1.04,~-2.62] \\ [-1.29,~-2.65] \end{pmatrix}$\\
Segue II &  149.4 & -38.1 &  35 & $ -39 $ & $ 58.7 $ & no PM &  [59,~527] & no prediction & no prediction \\
Willman 1 &  158.6 & 56.8 &  38 & $ -12 $ & $ 39.1 $ & no PM &  [50,~515] & $\begin{pmatrix} [-0.41,~+0.66] \\ [-1.12,~+1.33] \end{pmatrix}$ & $\begin{pmatrix} [-0.56,~-1.63] \\ [-1.44,~-3.88] \end{pmatrix}$\\
Coma Berenices &  241.9 & 83.6 &  44 & $  98 $ & $ 9.6 $ & no PM &  [96,~507] & $\begin{pmatrix} [-0.55,~+0.88] \\ [-0.79,~+0.85] \end{pmatrix}$ & $\begin{pmatrix} [-0.87,~-2.30] \\ [-1.16,~-2.79] \end{pmatrix}$\\
Bootes III &  35.4 & 75.4 &  47 & $ 198 $ & $ 2.8 $ & no PM &  [292,~480] & $\begin{pmatrix} [-0.30,~+0.61] \\ [-0.25,~+0.45] \end{pmatrix}$ & $\begin{pmatrix} [-1.42,~-2.33] \\ [-1.11,~-1.81] \end{pmatrix}$\\
LMC &  280.5 & -32.9 &  51 & $ 262 $ & $ 6.7 $ & $ 6.8_{-0.1}^{+0.5} $ &  [63,~482] & $\begin{pmatrix} [+0.70,~+2.55] \\ [+0.18,~+0.29] \end{pmatrix}$ & $\begin{pmatrix} [+0.42,~-1.43] \\ [+0.16,~+0.05] \end{pmatrix}$\\
SMC &  302.8 & -44.3 &  64 & $ 146 $ & $ 20.2 $ & $ 23.1_{-1.9}^{+2.7} $ &  [50,~461] & $\begin{pmatrix} [+0.57,~+1.36] \\ [-0.59,~-1.69] \end{pmatrix}$ & $\begin{pmatrix} [+0.39,~-0.40] \\ [-0.32,~+0.79] \end{pmatrix}$\\
Bootes &  358.1 & 69.6 &  66 & $  99 $ & $ 16.1 $ & no PM &  [128,~466] & $\begin{pmatrix} [-0.38,~+0.55] \\ [-0.43,~+0.35] \end{pmatrix}$ & $\begin{pmatrix} [-0.72,~-1.66] \\ [-0.71,~-1.50] \end{pmatrix}$\\
Draco &  86.4 & 34.7 &  76 & $ -291 $ & $ 10.4 $ & $ 24.2_{-4.0}^{+4.2} $ &  [96,~435] & $\begin{pmatrix} [-0.24,~+0.80] \\ [-0.07,~-0.12] \end{pmatrix}$ & $\begin{pmatrix} [-0.53,~-1.57] \\ [-0.05,~+0.00] \end{pmatrix}$\\
Ursa Minor &  105.0 & 44.8 &  76 & $ -247 $ & $ 21.7 $ & $ 25.9_{-3.4}^{+5.2} $ &  [86,~432] & $\begin{pmatrix} [-0.34,~+0.62] \\ [-0.14,~+0.29] \end{pmatrix}$ & $\begin{pmatrix} [-0.58,~-1.55] \\ [-0.24,~-0.67] \end{pmatrix}$\\
Sculptor &  287.5 & -83.2 &  86 & $ 111 $ & $ 5.0 $ & $ 160.2_{-12.1}^{+11.4} $ &  [91,~418] & $\begin{pmatrix} [+0.46,~+1.09] \\ [-0.57,~-1.19] \end{pmatrix}$ & $\begin{pmatrix} [+0.29,~-0.34] \\ [-0.39,~+0.24] \end{pmatrix}$\\
Sextans &  243.5 & 42.3 &  86 & $ 224 $ & $ 14.7 $ & $ 50.3_{-35.5}^{+45.3} $ &  [71,~418] & $\begin{pmatrix} [-0.09,~+0.47] \\ [-0.36,~+0.37] \end{pmatrix}$ & $\begin{pmatrix} [-0.21,~-0.77] \\ [-0.51,~-1.25] \end{pmatrix}$\\
Ursa Major &  159.4 & 54.4 &  97 & $ -55 $ & $ 36.0 $ & no PM &  [50,~406] & $\begin{pmatrix} [-0.12,~+0.14] \\ [-0.41,~+0.32] \end{pmatrix}$ & $\begin{pmatrix} [-0.19,~-0.46] \\ [-0.62,~-1.35] \end{pmatrix}$\\
Carina &  260.1 & -22.2 & 105 & $ 223 $ & $ 4.7 $ & $ 59.1_{-30.5}^{+30.6} $ &  [50,~399] & $\begin{pmatrix} [+0.28,~+0.94] \\ [+0.02,~+0.25] \end{pmatrix}$ & $\begin{pmatrix} [+0.09,~-0.57] \\ [-0.05,~-0.28] \end{pmatrix}$\\
Hercules &  28.7 & 36.9 & 132 & $  45 $ & $ 37.7 $ & no PM &  [185,~350] & $\begin{pmatrix} [-0.09,~+0.24] \\ [-0.23,~-0.17] \end{pmatrix}$ & $\begin{pmatrix} [-0.41,~-0.74] \\ [-0.28,~-0.34] \end{pmatrix}$\\
Fornax &  237.1 & -65.7 & 147 & $  55 $ & $ 14.6 $ & $ 15.0_{-0.4}^{+3.4} $ &  [50,~338] & $\begin{pmatrix} [+0.28,~+0.66] \\ [-0.25,~-0.47] \end{pmatrix}$ & $\begin{pmatrix} [+0.21,~-0.18] \\ [-0.21,~+0.00] \end{pmatrix}$\\
Leo IV &  264.4 & 57.4 & 154 & $ 132 $ & $ 2.0 $ & no PM &  [50,~335] & $\begin{pmatrix} [-0.11,~+0.12] \\ [-0.20,~+0.11] \end{pmatrix}$ & $\begin{pmatrix} [-0.19,~-0.42] \\ [-0.31,~-0.62] \end{pmatrix}$\\
Canes Venatici II &  113.6 & 82.7 & 160 & $ -129 $ & $ 4.2 $ & no PM &  [112,~327] & $\begin{pmatrix} [-0.13,~+0.10] \\ [-0.21,~+0.03] \end{pmatrix}$ & $\begin{pmatrix} [-0.25,~-0.48] \\ [-0.33,~-0.56] \end{pmatrix}$\\
Leo V &  261.9 & 58.5 & 178 & $ 173 $ & $ 1.1 $ & no PM &  [59,~306] & $\begin{pmatrix} [-0.11,~+0.08] \\ [-0.19,~+0.06] \end{pmatrix}$ & $\begin{pmatrix} [-0.15,~-0.34] \\ [-0.26,~-0.51] \end{pmatrix}$\\
Canes Venatici &  74.3 & 79.8 & 218 & $  31 $ & $ 1.5 $ & no PM &  [98,~262] & $\begin{pmatrix} [-0.12,~+0.03] \\ [-0.15,~-0.02] \end{pmatrix}$ & $\begin{pmatrix} [-0.19,~-0.34] \\ [-0.22,~-0.34] \end{pmatrix}$\\
Leo II &  220.2 & 67.2 & 233 & $  78 $ & $ 13.4 $ & $ 19.1_{-5.7}^{+26.8} $ &  [50,~245] & $\begin{pmatrix} [-0.06,~+0.04] \\ [-0.17,~-0.02] \end{pmatrix}$ & $\begin{pmatrix} [-0.11,~-0.21] \\ [-0.24,~-0.38] \end{pmatrix}$\\
Leo I &  226.0 & 49.1 & 254 & $ 282 $ & $ 20.4 $ & $ 86.3_{-19.1}^{+19.1} $ &  [198,~231] & $\begin{pmatrix} [+0.00,~+0.03] \\ [-0.10,~-0.07] \end{pmatrix}$ & $\begin{pmatrix} [-0.10,~-0.12] \\ [-0.25,~-0.29] \end{pmatrix}$\\

  \hline
 \end{tabular}
 \end{center}
 \small \smallskip
$l$, $b$: Heliocentric position of the MW satellite galaxies in Galactic longitude and latitude.
$r_{\sun}$: heliocentric distance.
$v_{\mathrm{los}}$: heliocentric line-of-sight velocity.
$\theta_{\mathrm{VPOS-3}}^{\mathrm{predicted}}$: The angle between the satellite galaxy position as seen from the MW centre and the VPOS-3 plane containing the MW centre, which is at the same time the smallest-possible inclination between the orbital plane of the satellite galaxy and the VPOS-3 plane.
$\theta_{\mathrm{VPOS-3}}^{\mathrm{measured}}$: The angle between the orbital plane of the satellite galaxy derived from its PM measurement (if available) with the orientation of the VPOS-3 plane.
$v_{\mathrm{min}}$, $v_{\mathrm{max}}$: adopted minimum and maximum absolute speeds of the satellite galaxies, used to constrain the range in PMs.
$\begin{pmatrix} \mu_{\alpha} \cos \delta \\ \mu_{\delta} \end{pmatrix}_{\mathrm{co}}$: predicted range in PMs assuming the satellite galaxy to co-orbit with respect to the majority of MW satellites with measured orbital poles.
 $\begin{pmatrix} \mu_{\alpha} \cos \delta \\ \mu_{\delta} \end{pmatrix}_{\mathrm{counter}}$: predicted range in PMs assuming the satellite galaxy to be counter-orbiting.
\end{minipage}
\end{table*}

\begin{figure*}
\centering
 \includegraphics[width=55mm]{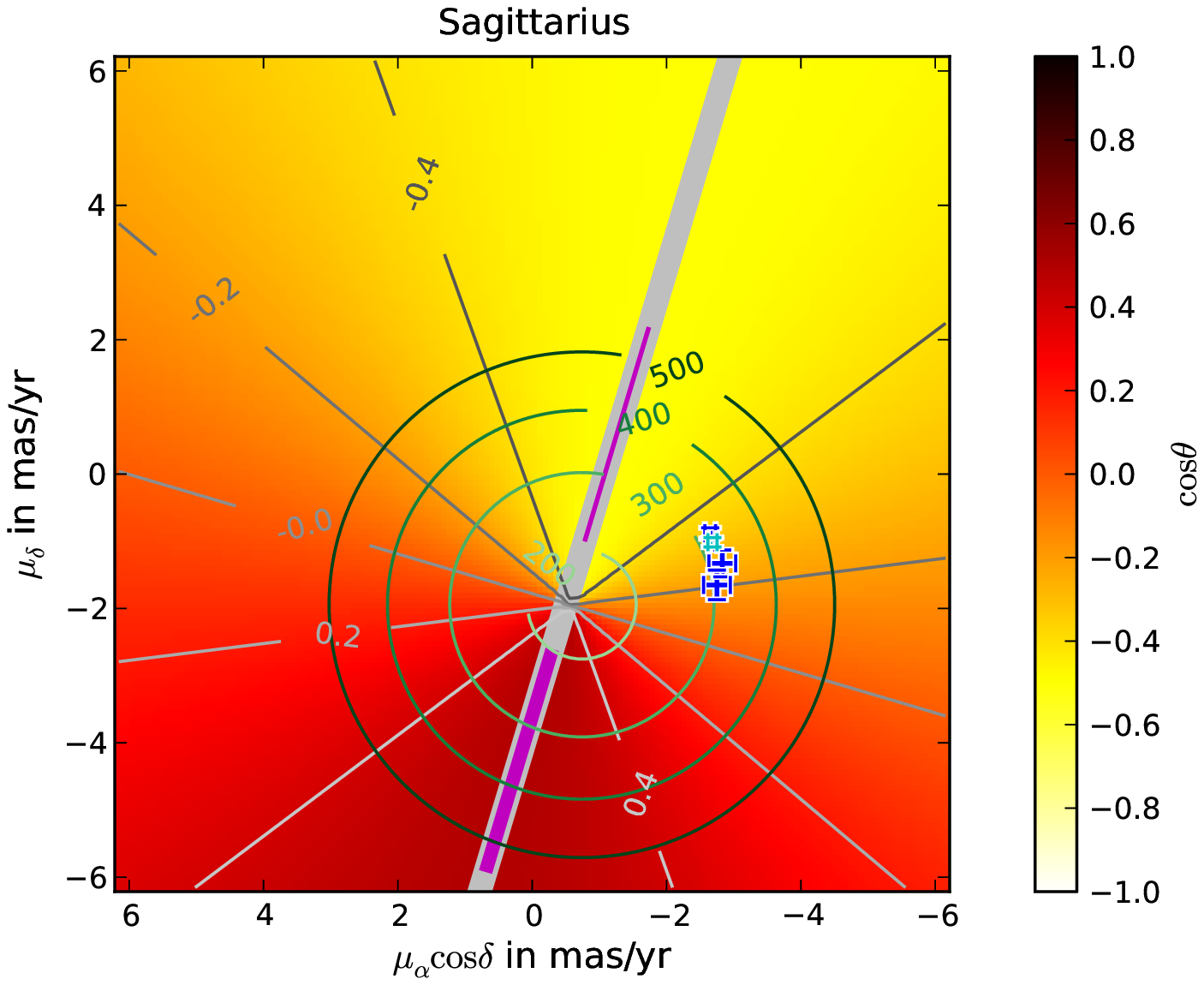}
 \includegraphics[width=55mm]{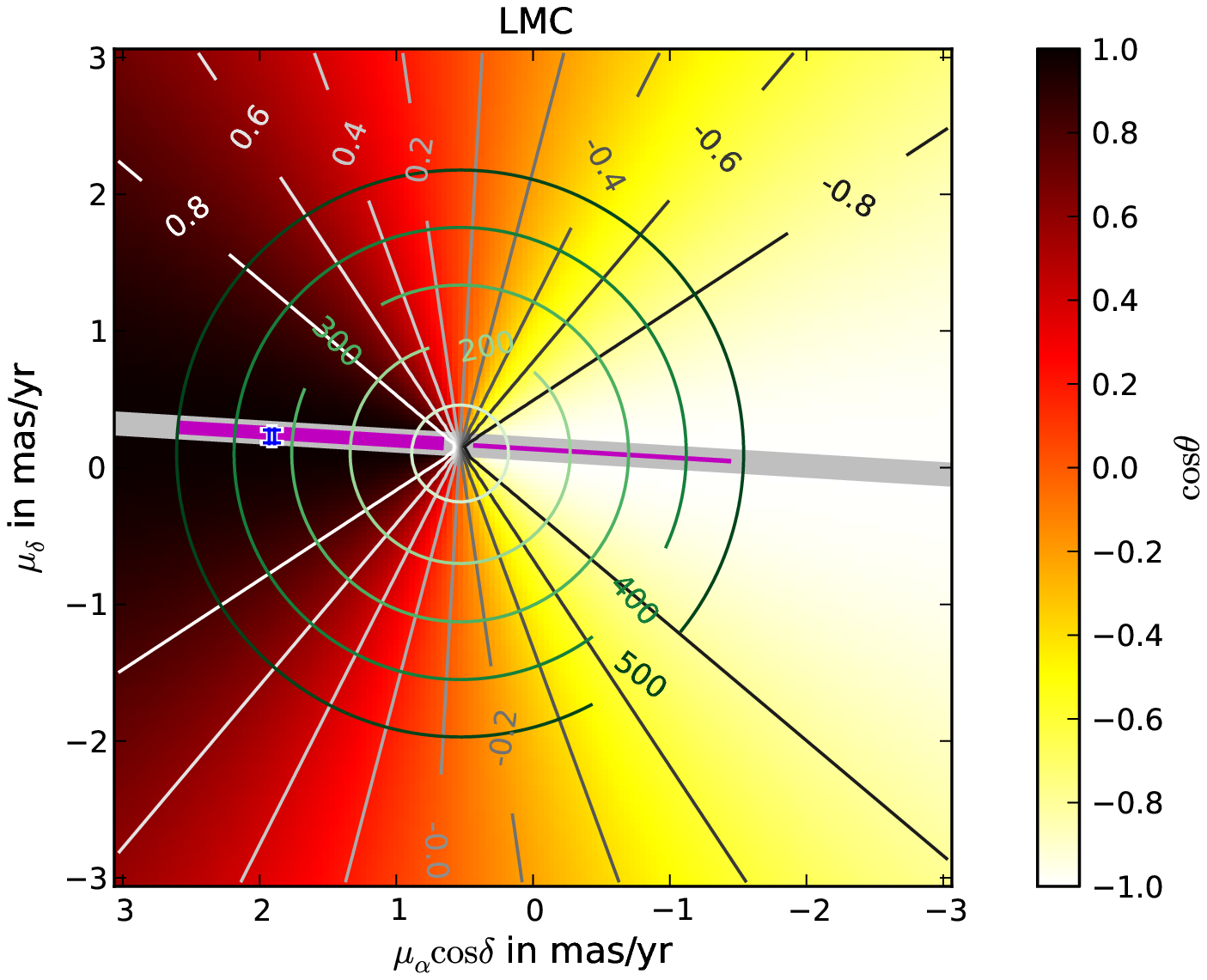}
 \includegraphics[width=55mm]{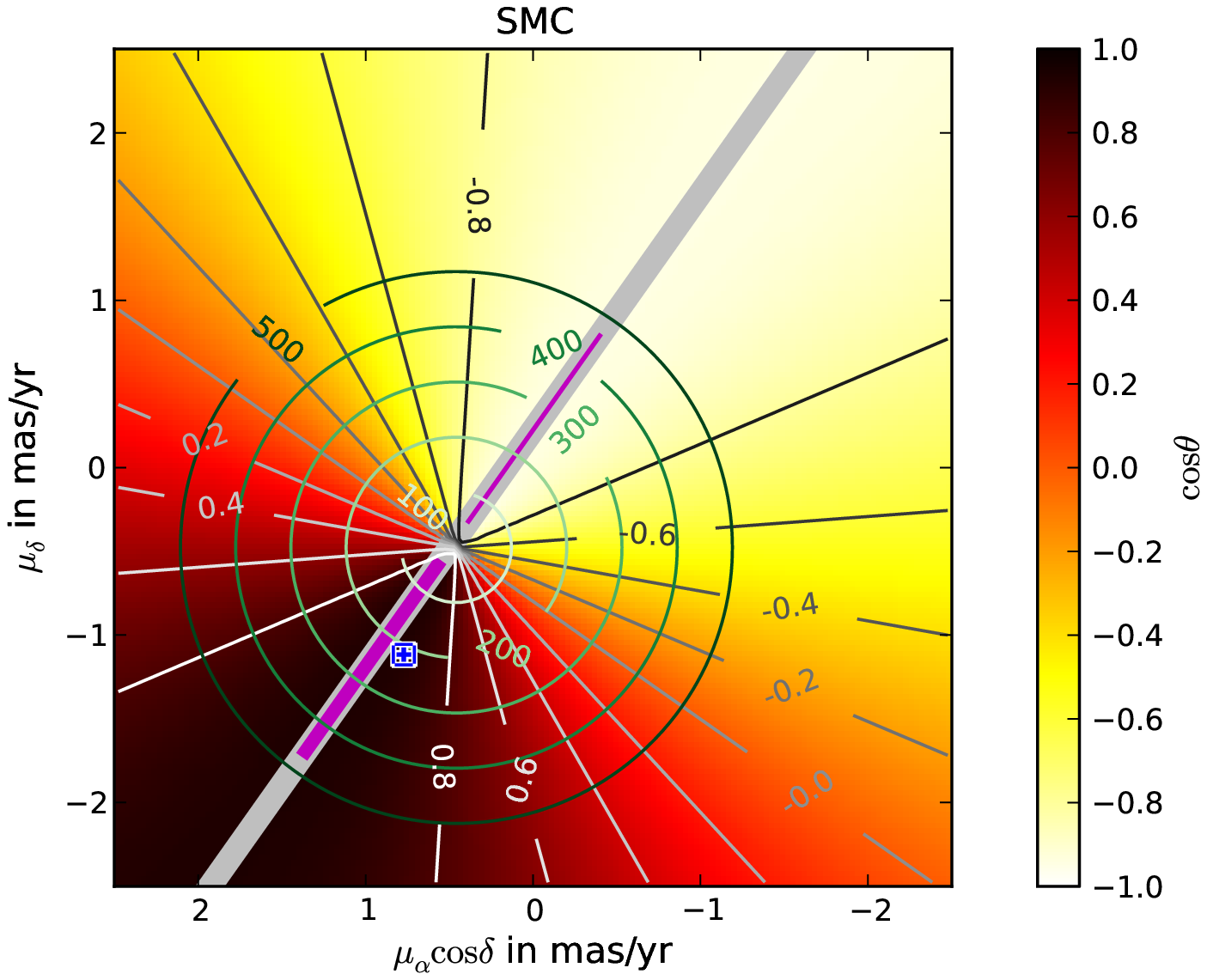}
 \includegraphics[width=55mm]{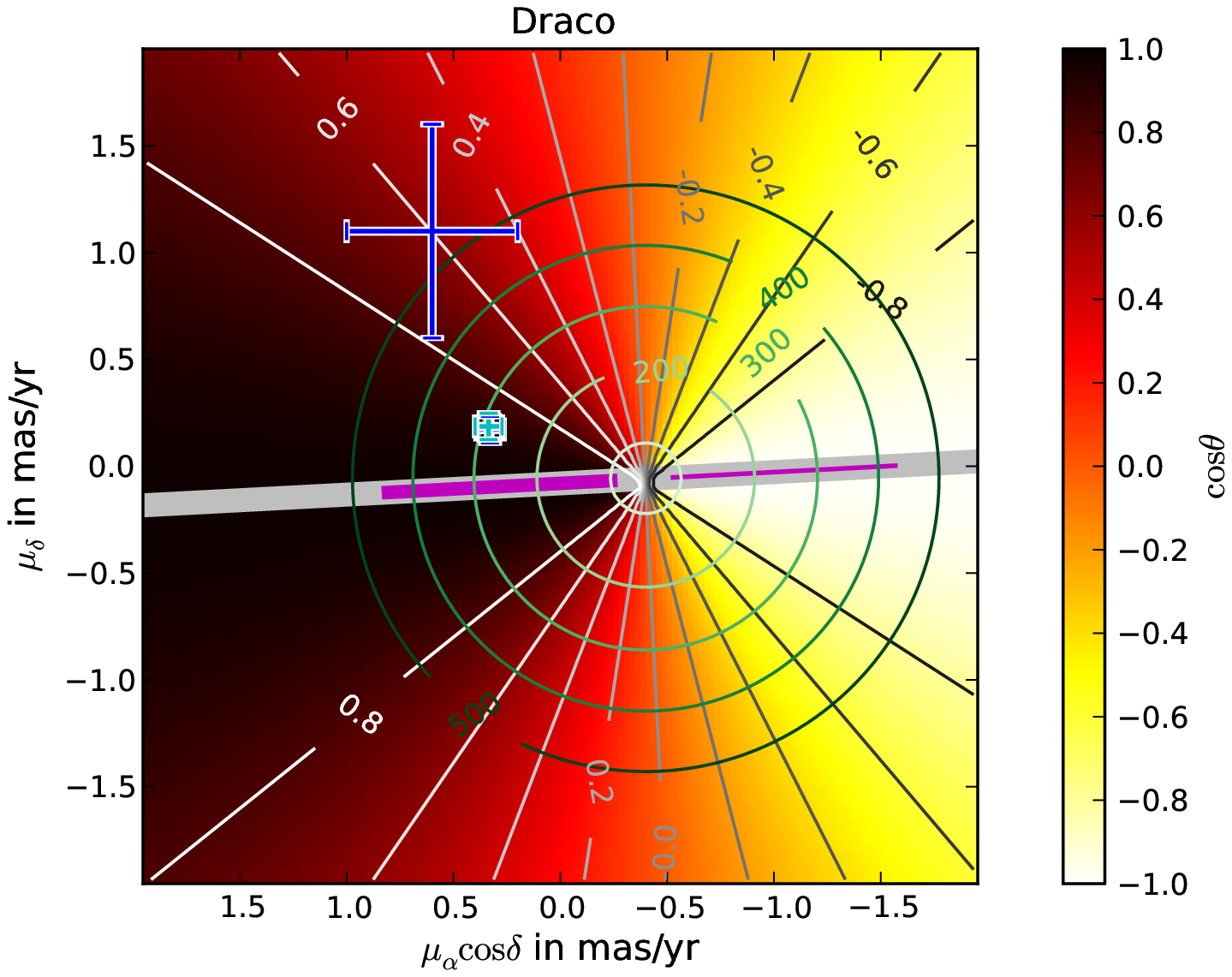}
 \includegraphics[width=55mm]{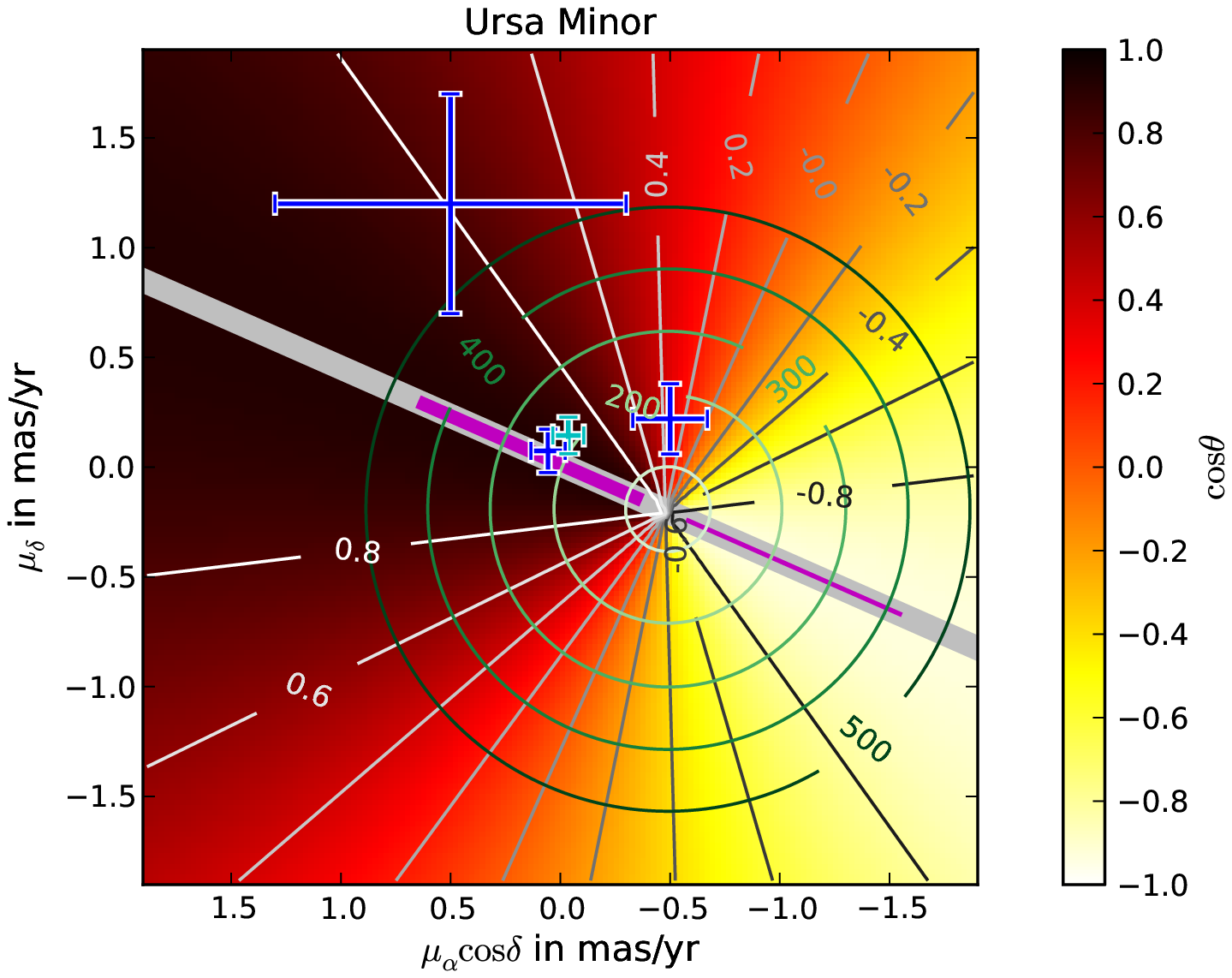}
 \includegraphics[width=55mm]{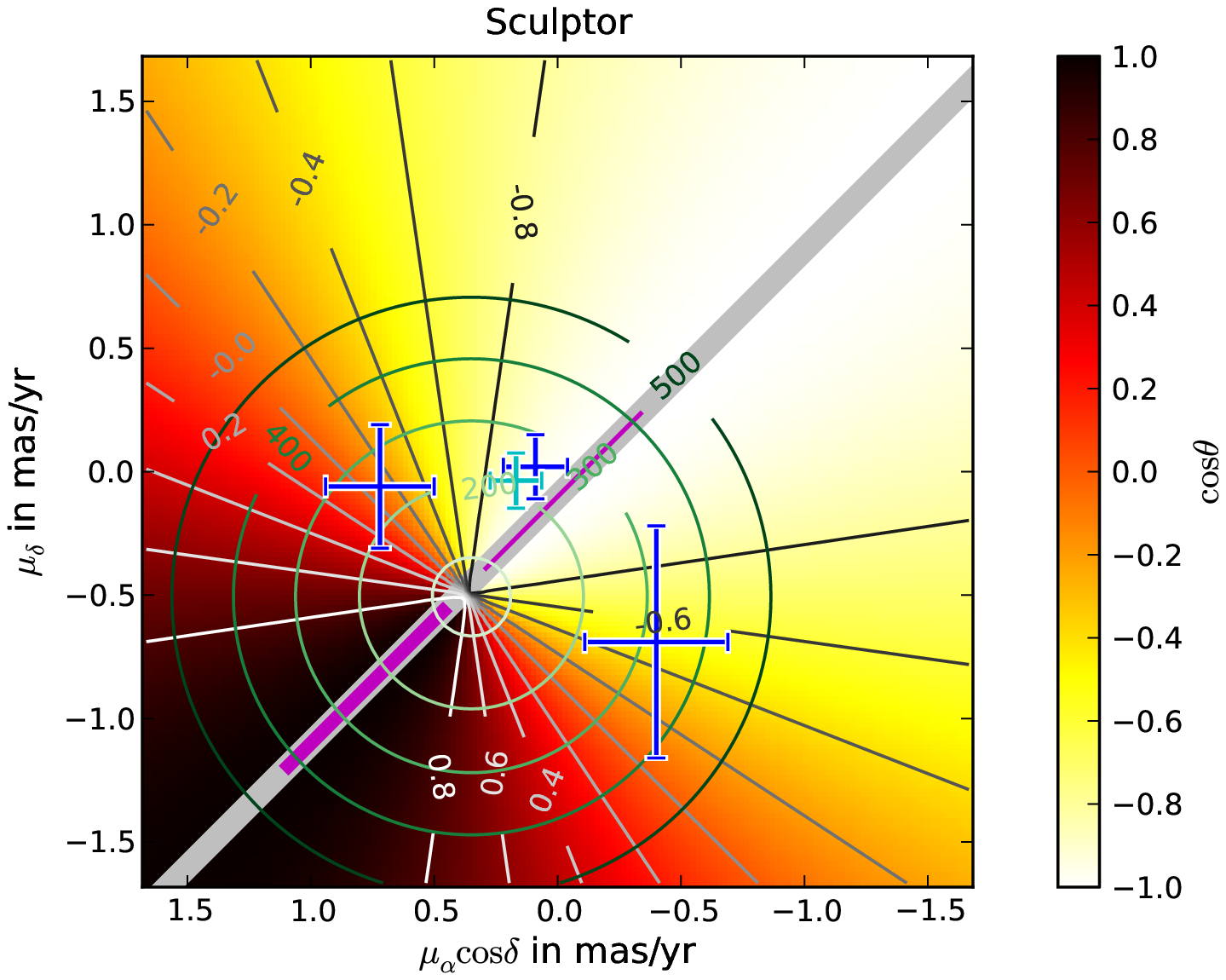}
 \includegraphics[width=55mm]{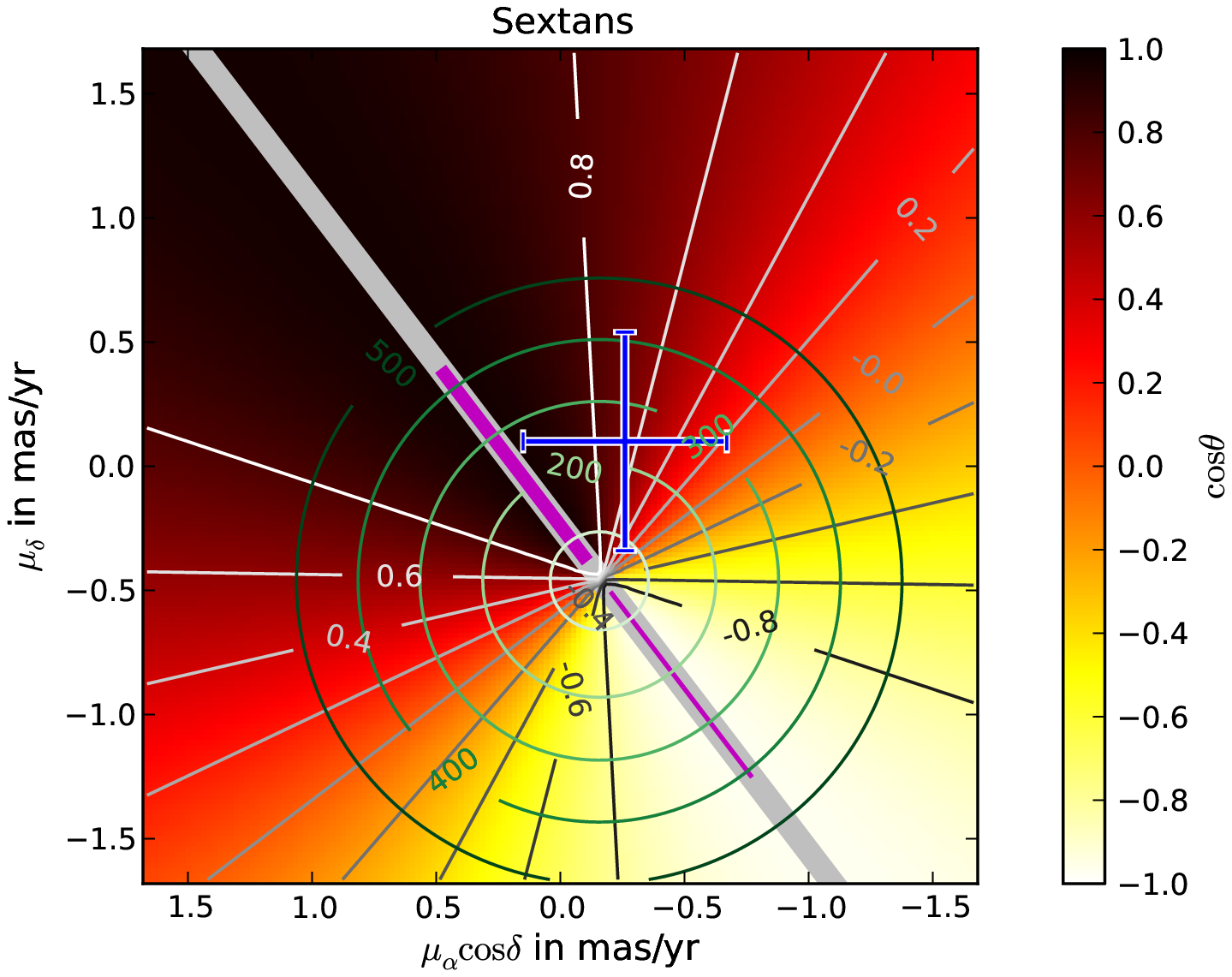}
 \includegraphics[width=55mm]{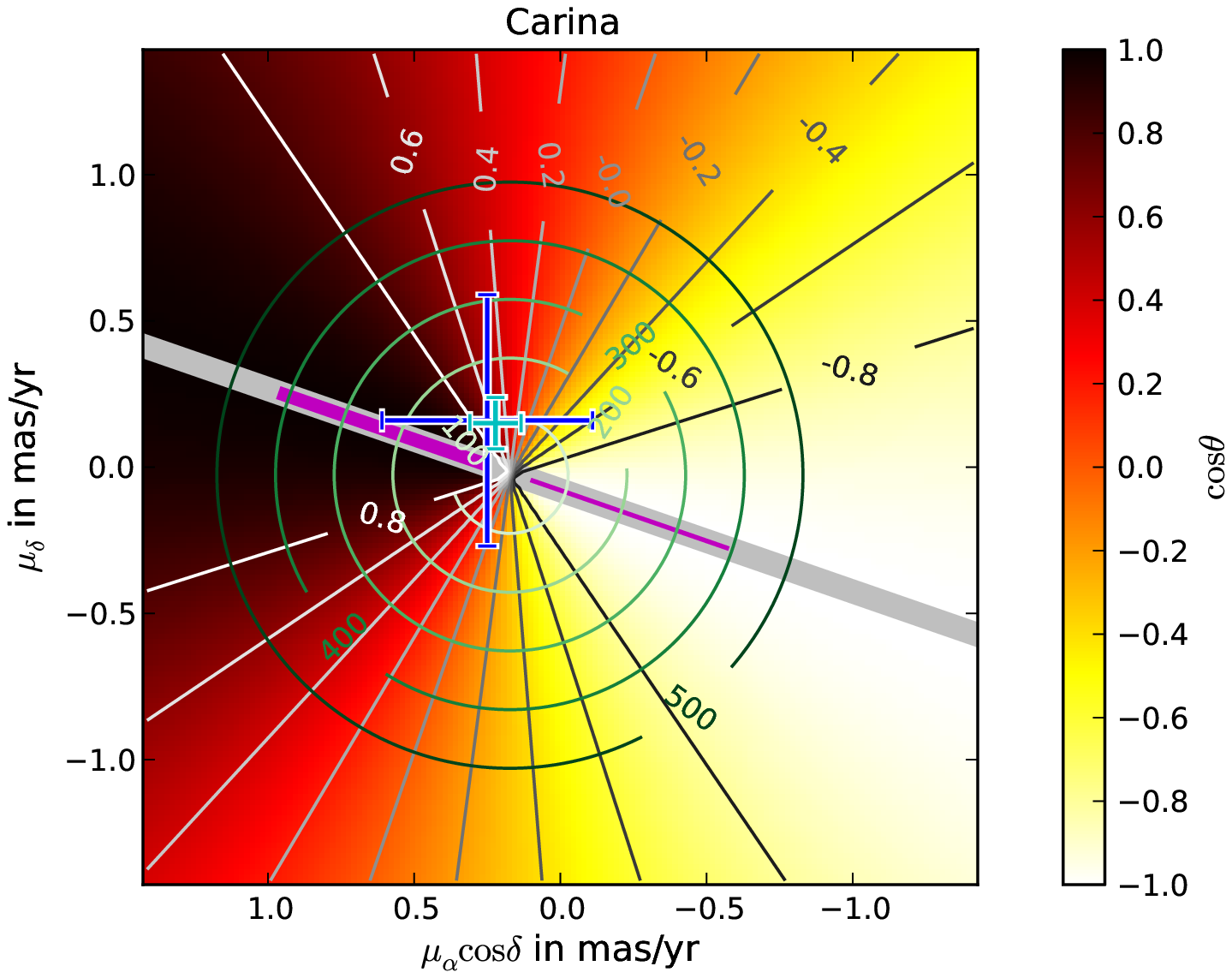}
 \includegraphics[width=55mm]{plots/predictPM_Fornax.eps}
 \includegraphics[width=55mm]{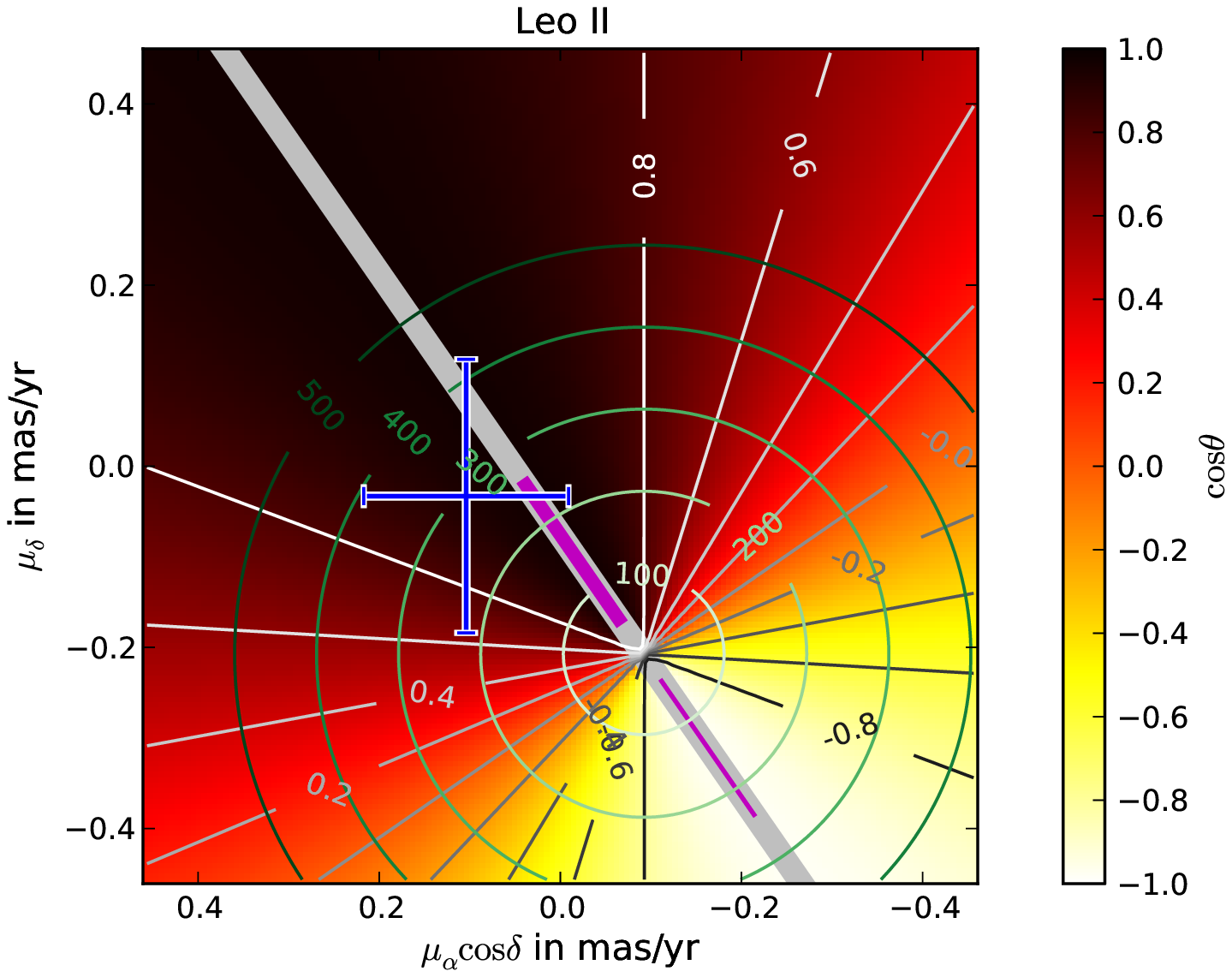}
 \includegraphics[width=55mm]{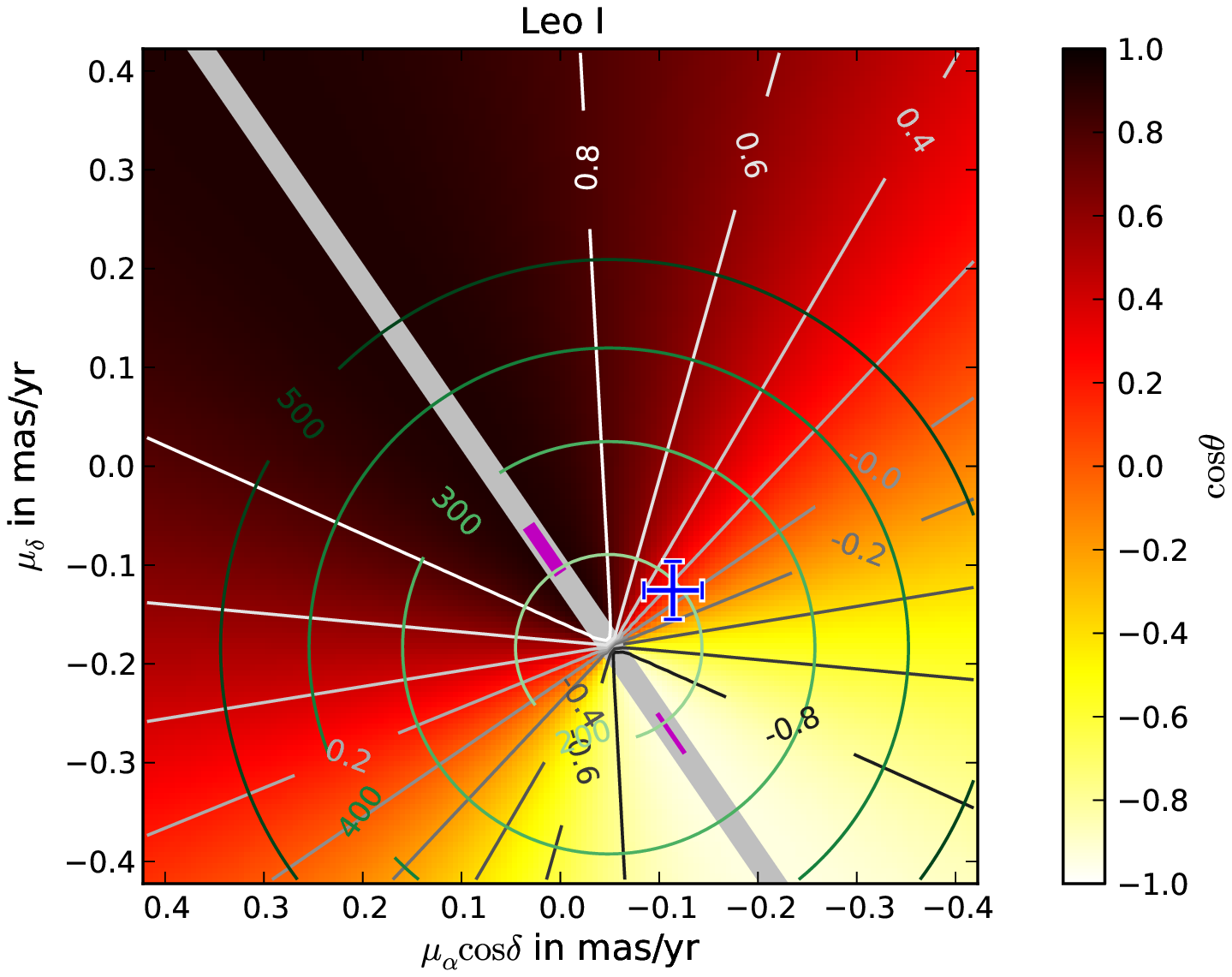}
 \caption{
 Predicted PMs for the MW satellites for which measured PMs are available. See Fig. \ref{fig:propmopredict} for a detailed discussion of the features in these plots. Note the low contrast in the first panel for Sagittarius. The position of this satellite galaxy does not allow it to orbit in a plane closely aligned with the VPOS-3. Therefore, no meaningful PM prediction is possible.
}
 \label{fig:propmopredictallclass}
\end{figure*}

\begin{figure*}
\centering
 \includegraphics[width=55mm]{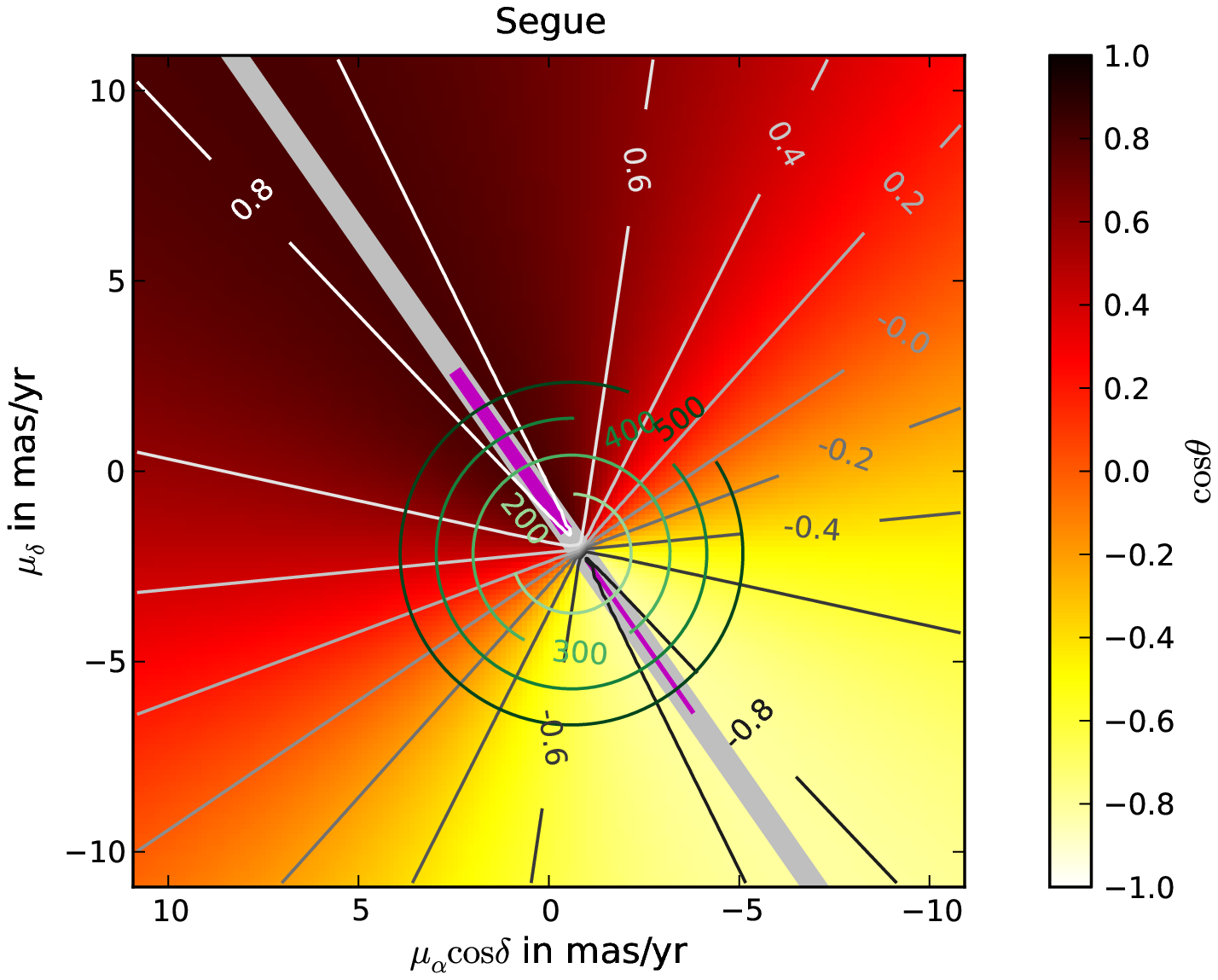}
 \includegraphics[width=55mm]{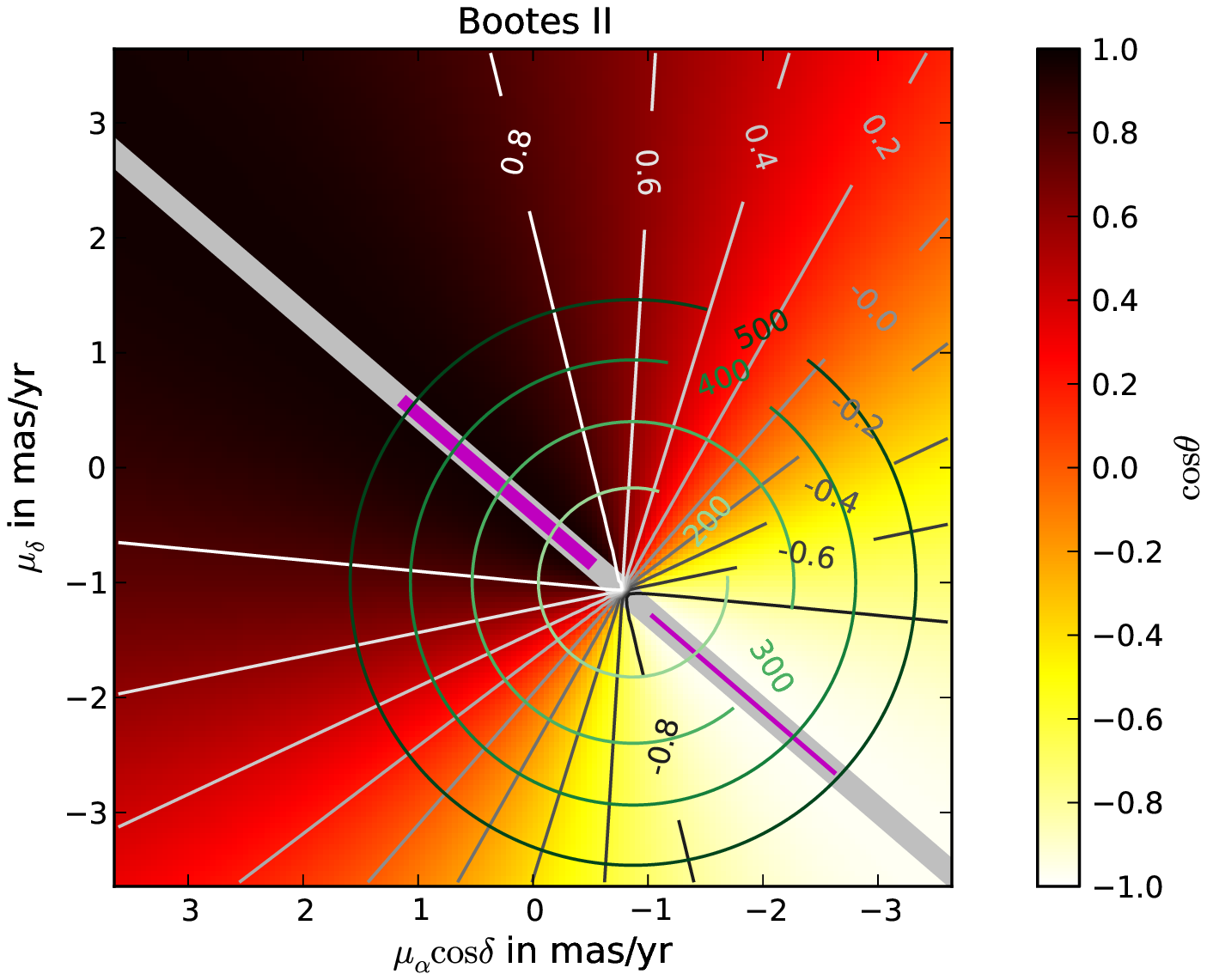}
 \includegraphics[width=55mm]{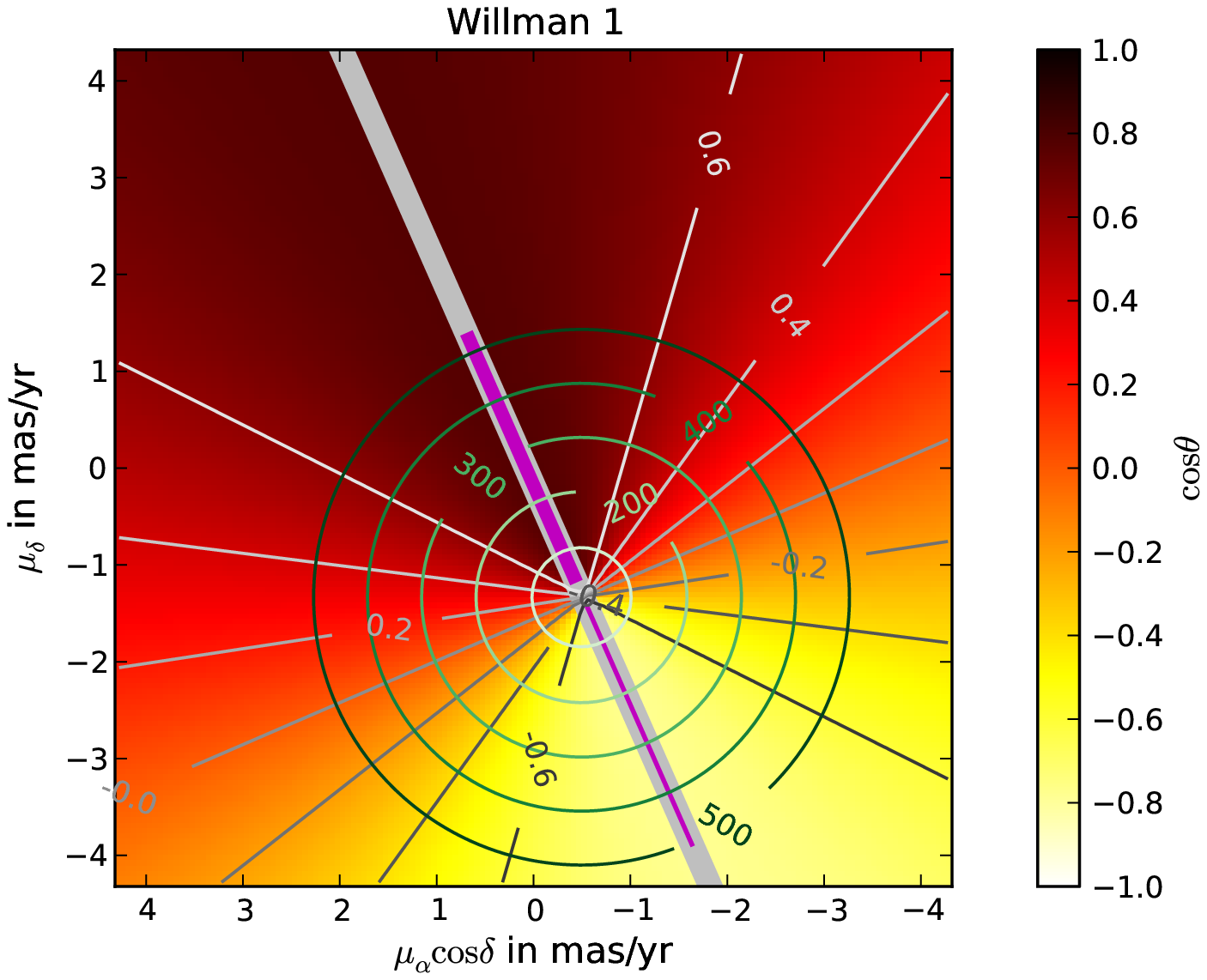}
 \includegraphics[width=55mm]{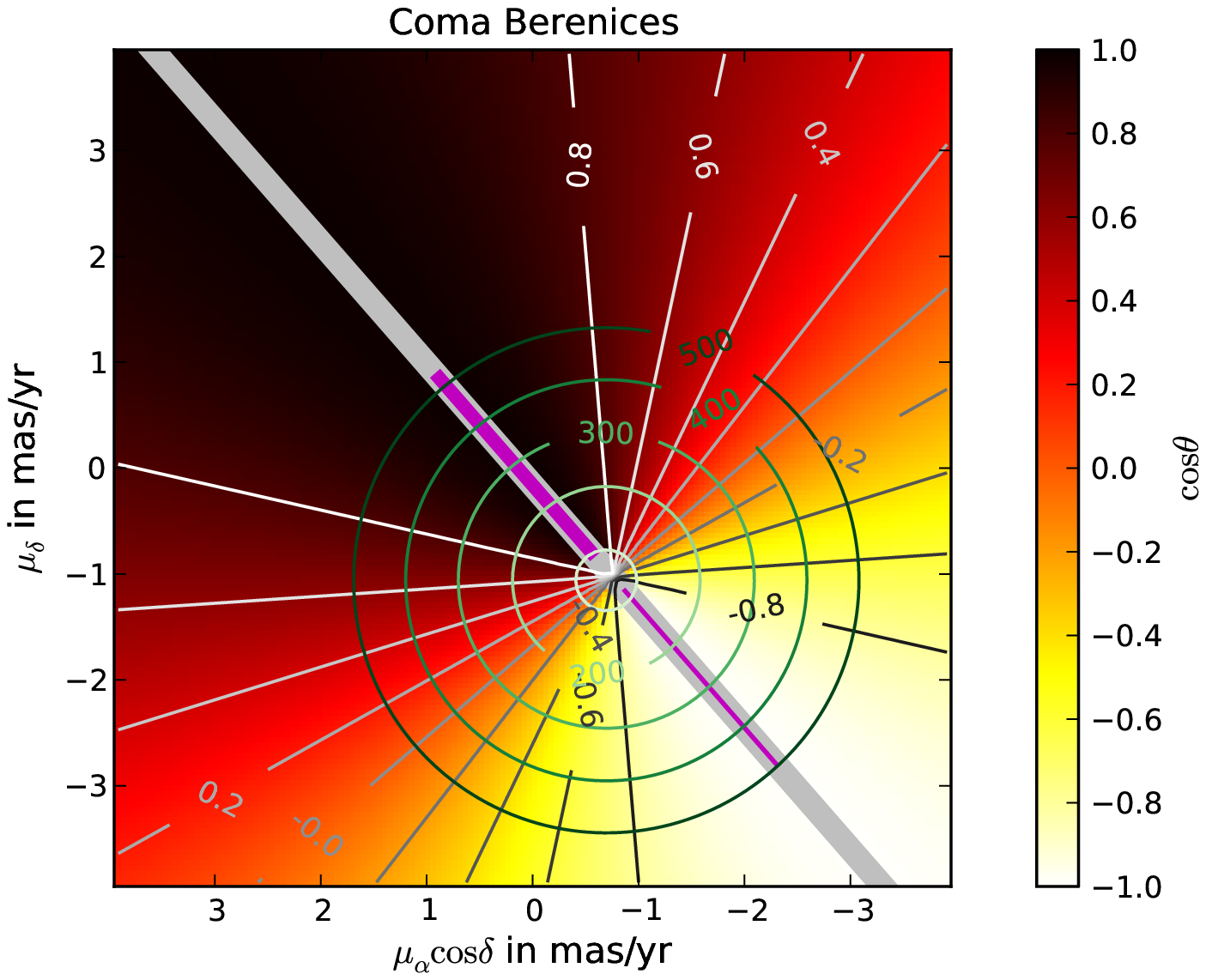}
 \includegraphics[width=55mm]{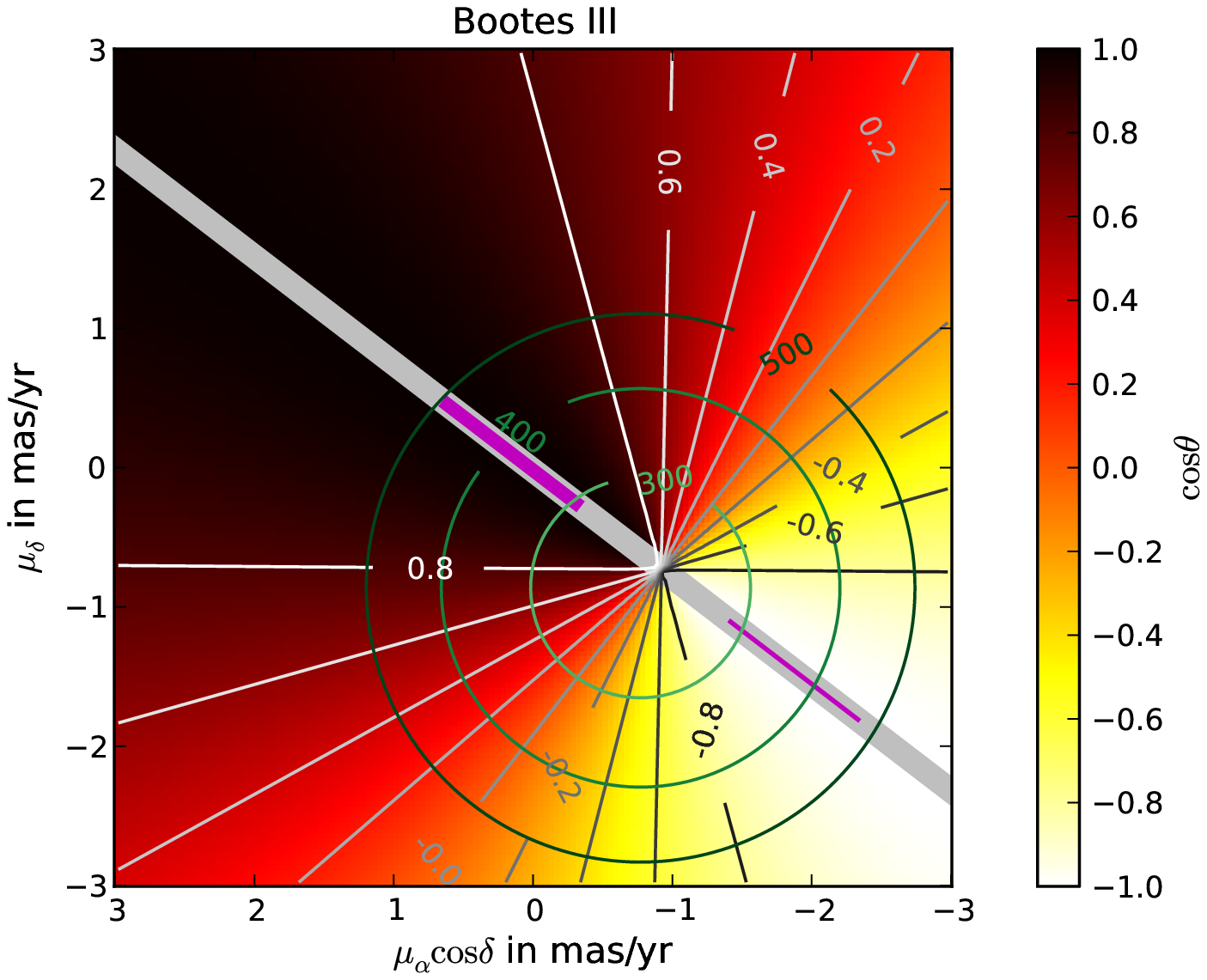}
 \includegraphics[width=55mm]{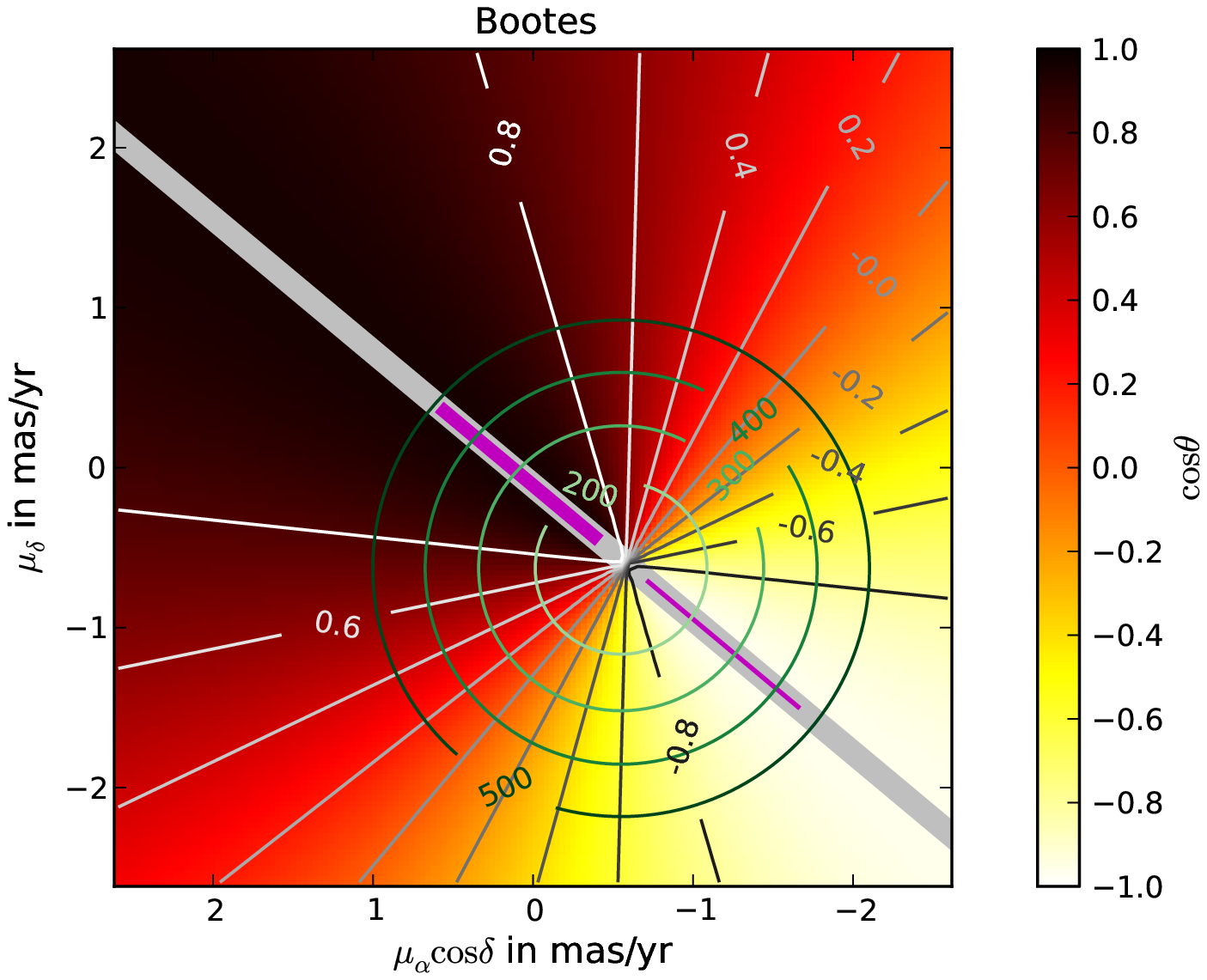}
 \includegraphics[width=55mm]{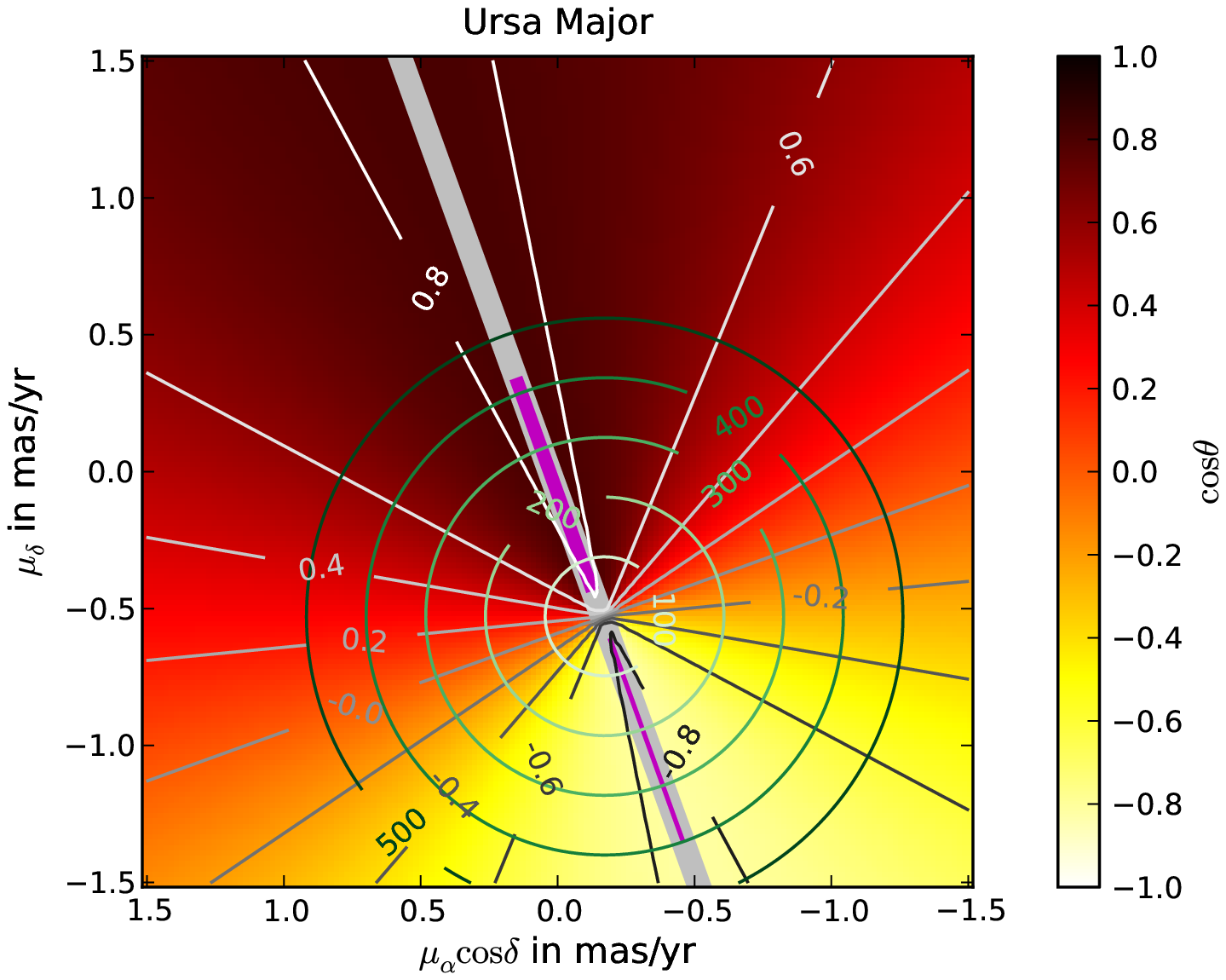}
 \includegraphics[width=55mm]{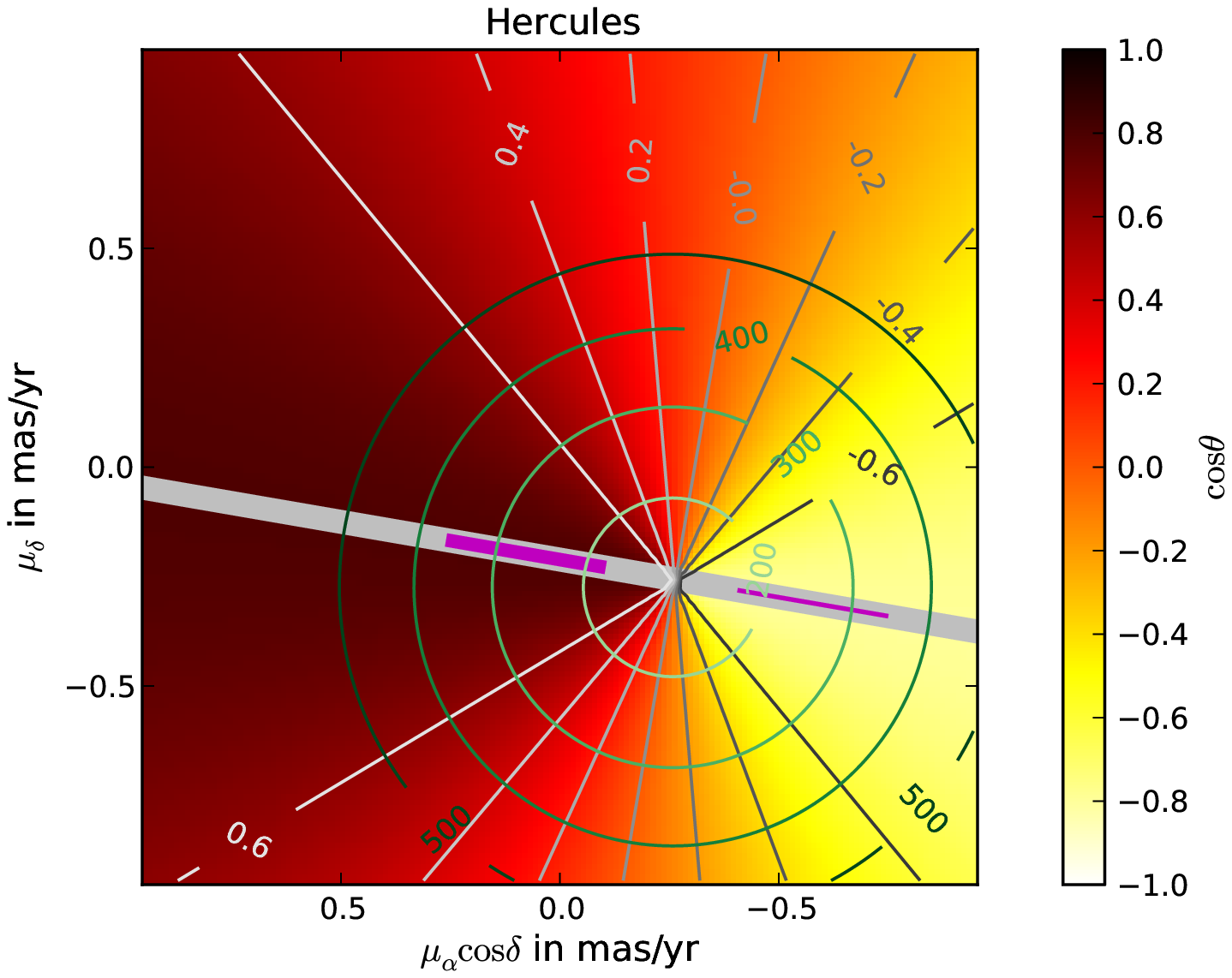}
 \includegraphics[width=55mm]{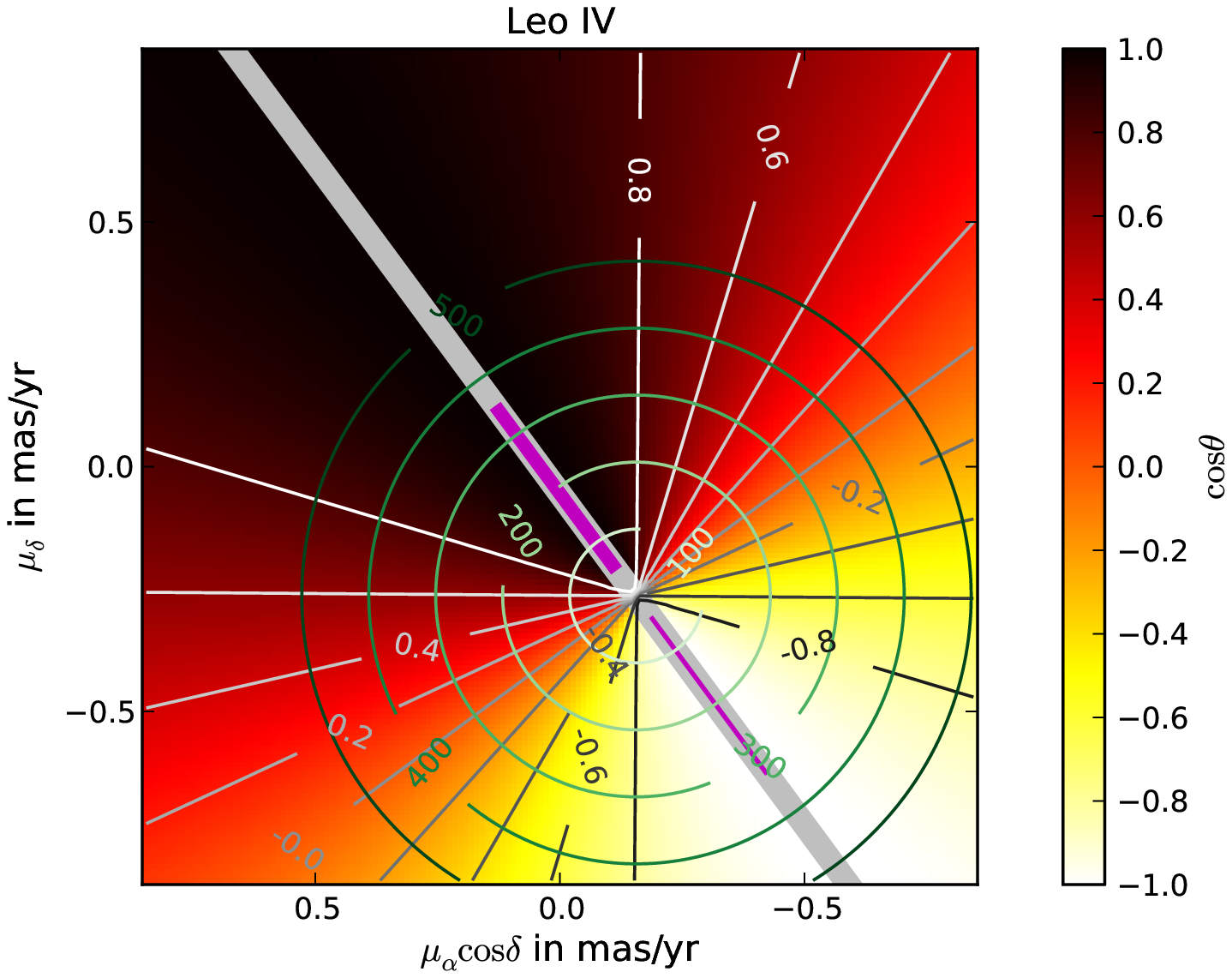}
 \includegraphics[width=55mm]{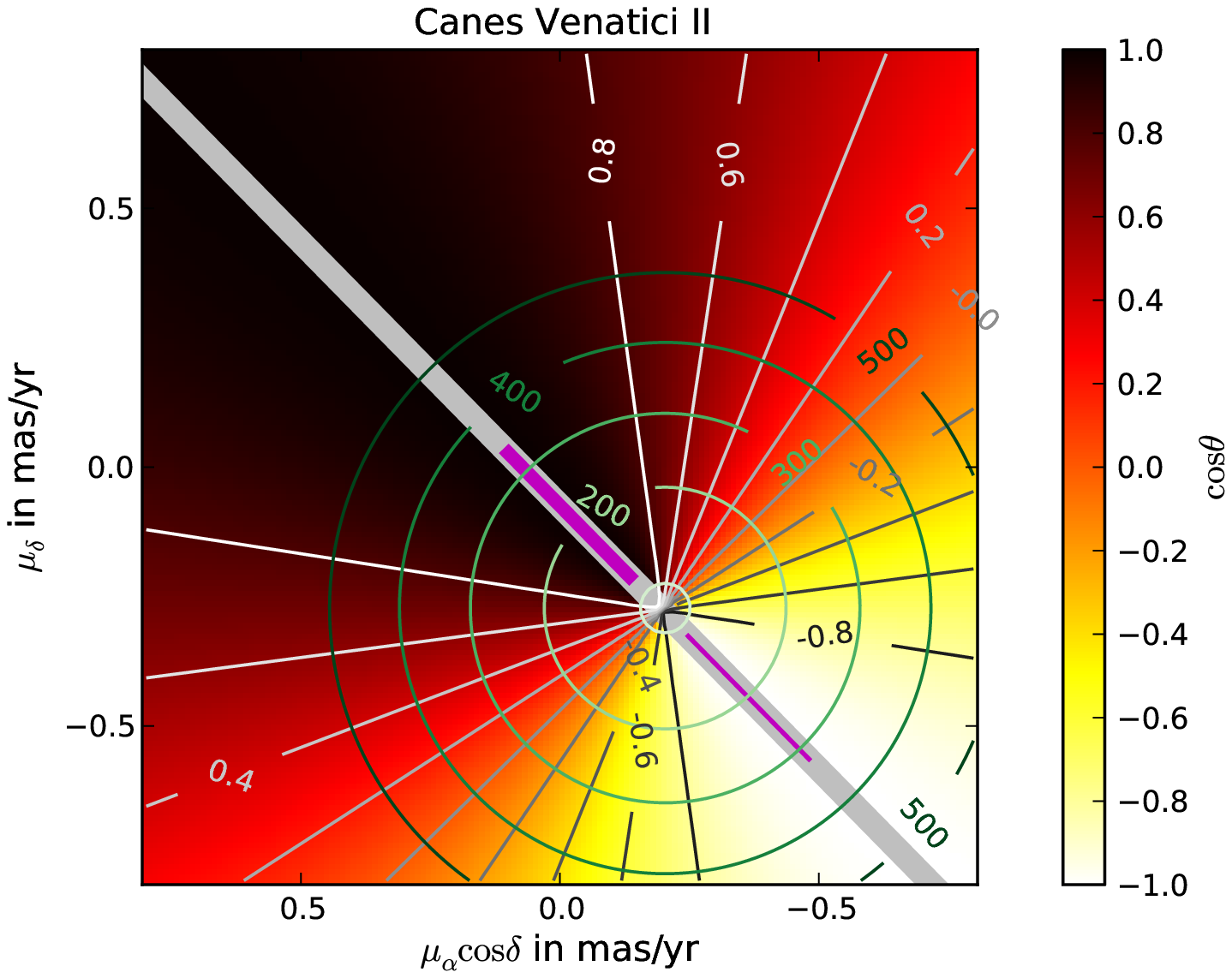}
 \includegraphics[width=55mm]{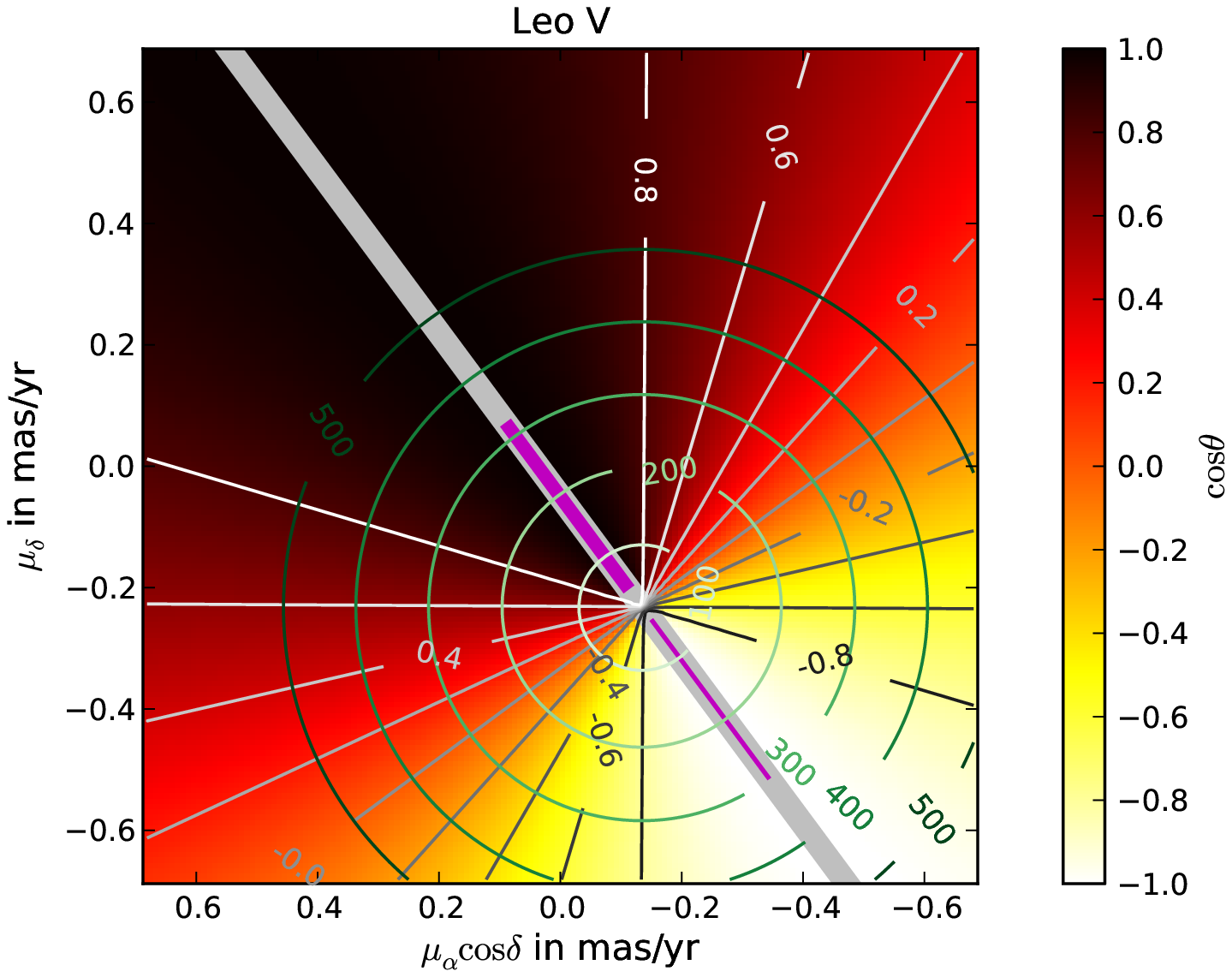}
 \includegraphics[width=55mm]{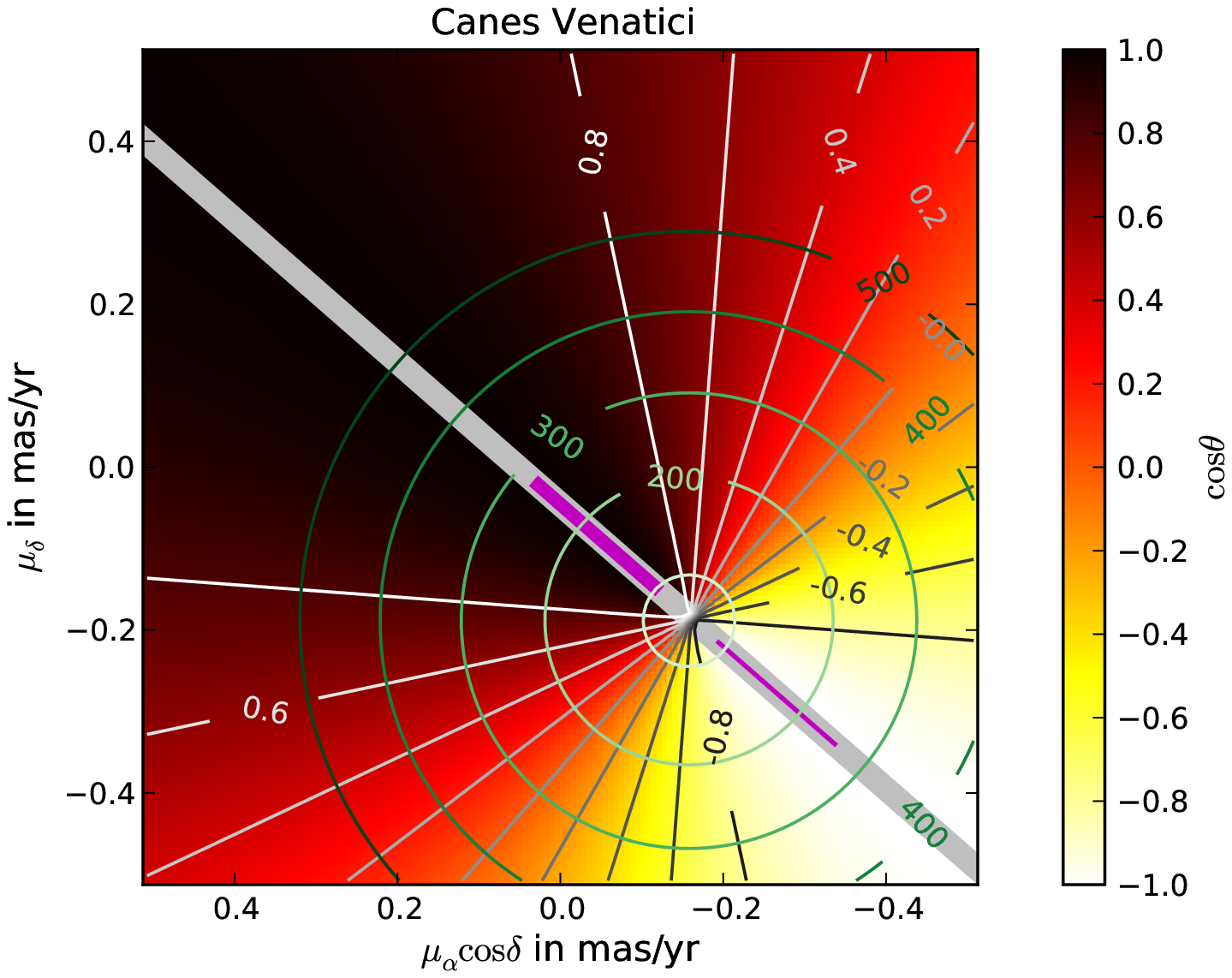}
 \caption{Predicted PMs for the MW satellite galaxies for which no measured PMs are available.}
 \label{fig:propmopredictallrest}
\end{figure*}

\begin{figure}
\centering
 \includegraphics[width=88mm]{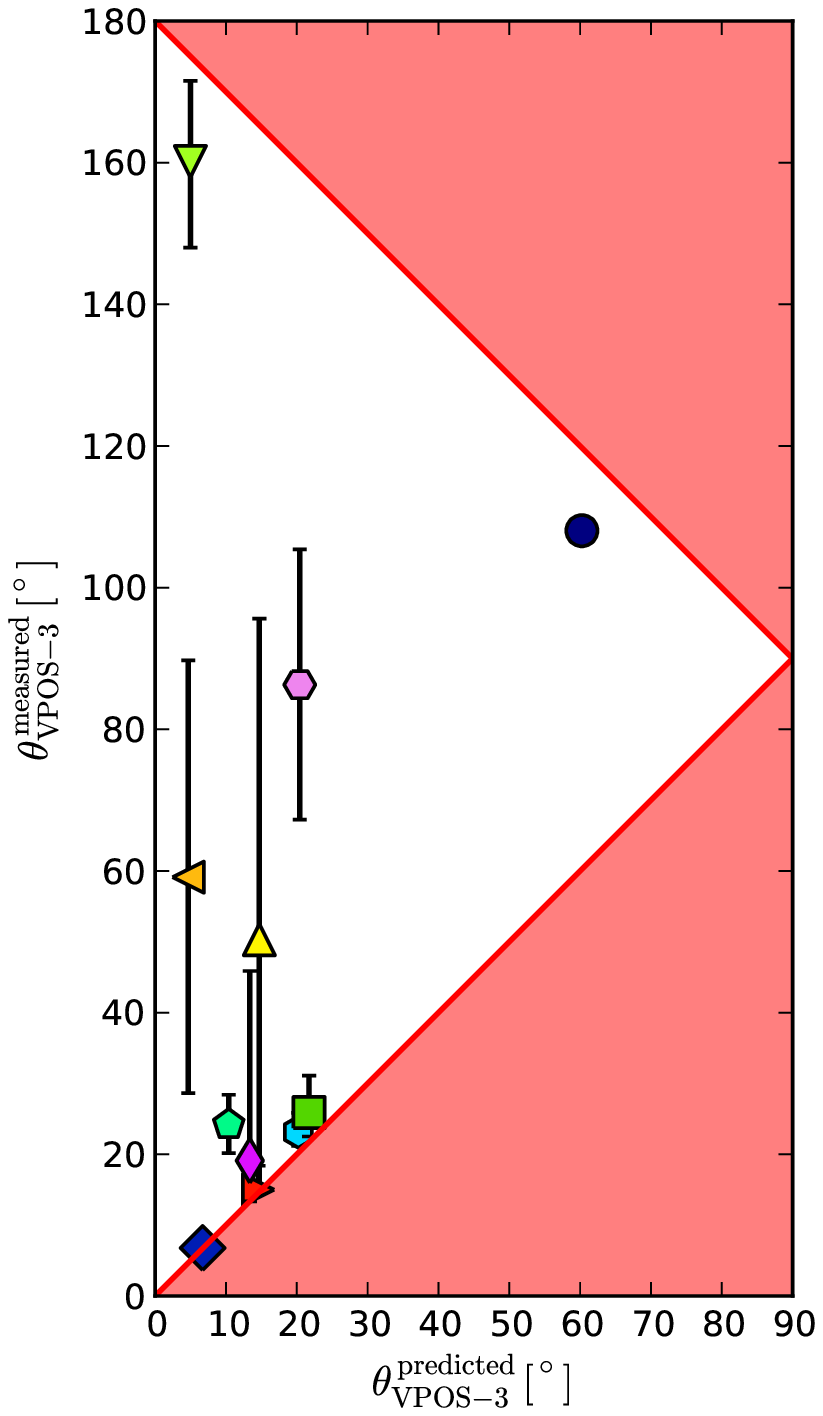}
 \caption{Comparison of predicted $\theta_{\mathrm{VPOS-3}}^{\mathrm{predicted}}$\ and measured $\theta_{\mathrm{VPOS-3}}^{\mathrm{measured}}$\ inclinations between the orbital planes of the MW satellites and the VPOS-3 plane. Same symbols as in Fig. \ref{fig:faceon}. The two red lines indicate the 1:1 relation between the predicted and measured values. If a prediction is met perfectly by the orbital pole derived from a satellite's PM, then it ends up on one of these lines. No measurement can end up in the shaded region because the predicted angle is always the minimum possible angle ($\theta_{\mathrm{VPOS-3}}^{\mathrm{predicted}} \leq \theta_{\mathrm{VPOS-3}}^{\mathrm{measured}}$). The upper half of the plot indicates counter-orbiting objects. It is particularly interesting that those satellite galaxies with the lowest uncertainties in their measured orbital pole direction (well-constrained PMs) are at the same time very close to the predicted values. This trend is expected if the satellites indeed orbit within the VPOS-3 because errors in the PM measurements can only result in a larger inclination from the true orbital plane.
 }
 \label{fig:comparepredictedmeasured}
\end{figure}

We have found that the majority of satellite galaxies for which PMs are available orbit close to a common plane, which is similar to the plane defined by their current positions. This indicates that the polar satellite galaxy structure is rotationally stabilized and that it will not disperse with time.

In addition to the 11 classical satellites for which PMs are available, the majority of the fainter satellites are also within the same planar structure, the VPOS around the MW \citep{Pawlowski2012a,Pawlowski2013a}. We can therefore assume that they, too, will orbit within this structure. This allows us to predict their expected orbital poles and therefore their PMs from their current positions and line-of-sight velocities. \textit{This prediction is model-independent and therefore empirical.} It only extrapolates the observationally found trend of well-concentrated orbital poles of the MW satellites to those objects for which no PM measurements are available yet. \textit{The predicted PMs do not depend on the exact formation scenario of the VPOS.}

The orientation of the plane close to which the satellites orbit allows us to predict the satellite orbital pole directions and thus their PMs. The orientation of this plane can be deduced from the current positions of the satellite galaxies because if their orbits are confined to a common plane their current positions will trace this plane. In \citet{Pawlowski2013a} it has been shown that the VPOS-3, a plane fitted to all MW satellites except for three outliers (Ursa Major, Hercules and Leo I), is a better representation of the satellite galaxy plane around the MW than the full VPOS fitted to all MW satellites including the outliers. The VPOS-3 orientation agrees with the peak in the density of plane normals fitted to all possible combinations of four MW satellites (see \citealt{Pawlowski2013a} for more details). The root-mean-square height of the MW satellites around the VPOS-3 is only 20\,kpc, in contrast to 29\,kpc for the full VPOS including only three additional objects. The normal direction of the VPOS-3 points to $(l,b) = (169^{\circ}.3, -2^{\circ}.8)$, such that it is very close to the normal of the Magellanic Stream and the LMC orbit (see Fig. \ref{fig:orbitalpolesASP}). It is also very well aligned with the average orbital pole of the most concentrated orbital poles discussed in the previous section. Finally, for one of the three outliers (Leo I) a PM measurement is available which results in an orbital plane that is strongly inclined to the orbital planes of the majority of the MW satellites (see the previous section). This further motivates removing at least Leo I from the plane fit for the prediction of the orbits. 

For the following prediction of the MW satellite PMs, we therefore adopt the VPOS-3 as the expected orientation of the average orbital plane of the satellite system. This will predict PMs which lead to a stable VPOS-3 in orientation and thickness by keeping the MW satellite orbital planes as close as possible to the VPOS-3 plane. The smallest possible angle between the orbital plane of a satellite galaxy and the VPOS-3, $\theta_{\mathrm{VPOS-3}}$, is defined by the current position of the satellite: it is simply the angle between the satellite galaxy position and the VPOS-3 plane as seen from the galactic centre. Satellite galaxies which currently have a large $\theta_{\mathrm{VPOS-3}}$\ will be unable to orbit within the VPOS-3, so for these no prediction is possible. We therefore refrain from predicting the PMs for those MW satellites for which $\theta_{\mathrm{VPOS-3}}$\ exceeds $45^{\circ}$. This is particularly true for the MW satellites close to the galactic centre, where additional effects such as the spatial thickness of the VPOS-3 come into play and where the MW potential is more strongly non-spherical, such that precession will influence the orientation of their orbits with time.

To predict the PMs, we proceed as follows. For each of the MW satellites, we collect the four known phase-space coordinates, which are its position in space and its heliocentric line-of-sight velocity (see first columns in Table \ref{tab:prediction}). We then scan a range of possible PMs. For each combination of $\mu_{\alpha} \cos \delta$\ and $\mu_{\delta}$\ we determine the direction of the orbital pole, keeping the four other known phase-space coordinates of the satellite galaxy fixed. We measure the inclination $\theta$\ of this possible orbital pole relative to the VPOS-3 plane. As an example, the resulting distribution of angles is plotted as a colour map in Fig. \ref{fig:propmopredict} for the satellite galaxy Fornax.

We then determine those PMs which minimize $\theta$\ to $\theta_{\mathrm{VPOS-3}}$. Because the absolute speed of the galaxy within its orbital plane does not affect the orbit's orientation, this results in a line in the $\mu_{\alpha} \cos \delta$ -- $\mu_{\delta}$\ plane along which the satellite galaxy would have different absolute speeds (grey line in Fig. \ref{fig:propmopredict}; the different absolute speeds are indicated by the circular contours). We constrain the minimum and maximum of the expected PM by defining a minimum and a maximum accepted absolute speed for each satellite galaxy relative to the MW. The minimum observed absolute speed of a galaxy is given by its radial velocity component, assuming its tangential velocity to be zero. We adopt 1.2 times this radial velocity (relative to the MW) as the minimum velocity $v_{\mathrm{min}}$\ to avoid predicting completely radial orbits, but demand a minimum absolute speed of $50\,\mathrm{km\,s}^{-1}$\ if the radial velocity is very low.

For the maximum speed, we assume that the MW potential can be approximated with a spherical logarithmic potential of the form
\begin{displaymath}
\Phi_{\mathrm{log}}(r) = \frac{1}{2} v_{\mathrm{circ}}^2 \ln\left(r^2 + r_{\mathrm{0}}^2\right) + \mathrm{const.},
\end{displaymath}
with a circular velocity $v_{\mathrm{circ}}$\ at a scale radius $r_0$. We set these values equal to the ones adopted for the MW in Sect. \ref{sect:datasets}: 
$r_0 = d_{\sun} = 8.3$\,kpc and $v_{\mathrm{circ}} = v_{\mathrm{LSR}} = 239\,\mathrm{km}\,\mathrm{s}^{-1}$\ \citep{McMillan2011}.
 We then determine the speed $v_{\mathrm{max}} (r)$\ a galaxy at a given radial distance $r$\ from the centre of the MW needs in order to be able to reach the saddle point in the potential of the MW and M31. For simplicity, we assume both galaxies to be of equal mass and separated by 800\,kpc. Therefore, a galaxy which is able to reach a Galactocentric distance of 400\,kpc can be considered an LG galaxy and not a bound satellite galaxy bound to the MW. This speed $v_{\mathrm{max}} (r)$\ is similar to an escape speed but more physically motivated regarding the LG geometry.
The effective potential for our assumption is therefore
\begin{displaymath}
\Phi_{\mathrm{eff}}(r) = \Phi_{\mathrm{log}}(400\,\mathrm{kpc}) - \Phi_{\mathrm{log}}(r),
\end{displaymath}
for $r \leq 400\,\mathrm{kpc}$. The maximum speed is calculated analogously to the escape speed from a potential
\begin{displaymath}
v_{\mathrm{max}} (r) = v_{\mathrm{esc}} (r) = \sqrt{2 |\Phi_{\mathrm{eff}}(r)|}.
\end{displaymath}
So for the assumed potential
\begin{displaymath}
v_{\mathrm{max}} (r) = \sqrt{v_{\mathrm{circ}}^2 \left( \ln\left((400\,\mathrm{kpc})^2 + r_0^2\right) - \ln\left(r^2 + r_0^2\right) \right) }
\end{displaymath}
\begin{displaymath}
= \sqrt{v_{\mathrm{circ}}^2 \ln \left( \frac{(400\,\mathrm{kpc})^2 + r_0^2}{r^2 + r_0^2} \right) },
\end{displaymath}
with $v_{\mathrm{circ}} = 239\,\mathrm{km}\,\mathrm{s}^{-1}$\ and $r_0 = d_{\sun} = 8.3$\,kpc. This function is plotted as a red line in Fig. \ref{fig:satvel}, together with the measured absolute speeds of the MW satellites for which PMs are available. With the exception of the two outermost MW satellites Leo I and Leo II, all satellites fall well below this speed.

The predicted PMs are therefore along the grey line in the $\mu_{\alpha} \cos \delta$ -- $\mu_{\delta}$\ plane and in between $v_{\mathrm{min}}$\ and $v_{\mathrm{max}}$\ (Fig. \ref{fig:propmopredict}). We also discriminate between the co- and the counter-orbiting direction of the orbit relative to the average observed orbital pole direction. Table \ref{tab:prediction} compiles the input parameters for all MW satellites \citep{McConnachie2012} and the resulting predictions.
The predicted ranges in PMs are also illustrated in Fig. \ref{fig:propmopredictallclass} for all MW satellites with available PM measurements and in Fig. \ref{fig:propmopredictallrest} for all remaining MW satellites for which meaningful predictions are possible.

Fig. \ref{fig:propmopredictallclass} allows us to judge the quality of the predictions. Many of the satellite galaxy PMs are indeed very close to the magenta line indicating the predicted range of PMs. This is particularly true for the LMC and SMC, which have tightly constrained PM measurements overlapping well with the predicted range.

Another way to demonstrate how well the prediction works is shown in Fig. \ref{fig:comparepredictedmeasured}, which compares the angle between the VPOS-3 and the predicted orbital plane of each satellite with the angle between the VPOS-3 and the measured orbital plane. Of the 11 satellites, only Sagittarius cannot be closely aligned with the VPOS-3 plane; the other satellites can have orbital planes inclined by only $\approx 20^{\circ}$\ or less from the VPOS-3. The majority of these indeed come close to the predicted inclinations in a co-orbiting sense (lower half of the plot); only Sculptor is counter-orbiting (upper half). Interestingly, those satellite galaxies which have the least uncertain orbital pole directions (small error bars) all cluster close to the red line indicating perfect agreement between the predicted and the measured orbital poles. In particular, Carina and Sextans are more distant than predicted, but within their very large uncertainties they can still align well with the VPOS-3.
Such a behaviour is to be expected if the satellite galaxies indeed orbit within the VPOS-3 plane. A large measurement uncertainty in the PM makes it more likely that the resulting measured orbital pole is inclined relative to the true orbital pole. As the measurement becomes more precise and the uncertainty decreases, the measured PM will approach the true one and the measured orbital pole will come closer to the true orbital pole.

\section{Conclusion}
\label{sect:conclusion}

We have collected the available PM measurements for the 11 brightest (classical) MW satellite galaxies (Table \ref{tab:satellitedata}). These were used to determine the orbital poles, the direction of the orbital angular momenta of the satellites around the MW and the orbital pole uncertainties (Fig. \ref{fig:orbitalpolesASP} and Table \ref{tab:satellitepoles}). Our results confirm the earlier finding that the MW satellite galaxies preferentially co-orbit in a similar direction \citep{Metz2008}, but the updated and extended PM list now puts this finding on to a solid ground. The orbital poles of at least six satellites (LMC, SMC, Ursa Minor, Leo II, Draco and Fornax) are closely aligned with each other, being inclined by less than $27^{\circ}$\ from their average direction (which points to $[l,b] = [177^{\circ},-3^{\circ}])$\ and close to the minor axis of the VPOS. Two additional satellite galaxies (Sextans and Carina) agree with this common direction within their uncertainties and one satellite galaxy (Sculptor) is counter-orbiting within the same orbital plane. Interestingly the two non-aligned satellites are the innermost and outermost MW satellite galaxies (Sagittarius and Leo I, respectively).

The VPOS around the MW is rotationally stabilized by virtue of the Galactocentric tangential velocity components being larger than the radial values (Fig. \ref{fig:satvel}).
The orbital pole of the LMC, as well as the normal to the plane containing the Magellanic Stream and the VPOS-3, the plane fitted to all known MW satellites except three outliers, all are very close to the centre of the concentrated orbital pole distribution. We can therefore predict the PMs of the remaining MW satellites which are close to the VPOS, by requiring them to orbit close to the VPOS-3 plane (see Table \ref{tab:prediction} and Figs. \ref{fig:propmopredictallclass} and \ref{fig:propmopredictallrest}). \textit{This prediction is entirely empirical and therefore model-independent.} 
As a test of our method, comparing the inclinations of the predicted with the measured orbital planes of the MW satellites demonstrates that those satellites with better-constrained PM closely follow the prediction (Fig. \ref{fig:comparepredictedmeasured}). 
Such a trend is expected if the prediction is valid because more uncertain measurements tend to deviate more from the true value.

The radial distribution of speeds of those satellite galaxies for which PM measurements exist appear to trace an effective potential with a constant circular speed of approximately $240\,\mathrm{km\,s}^{-1}$\ to a Galactocentric distance of 250\,kpc. This corresponds to the circular speed of a logarithmic (phantom, \citealt{Famaey2012}) dark matter halo of the MW. 

Comparisons with the expectations derived from cosmological simulations demonstrate that the observed orbital pole distribution is extremely unlikely if the MW satellite galaxies are sub-haloes in a $\Lambda$CDM universe (see Sect. \ref{subsect:discussion2}). Within the observed orbital pole uncertainties it is by now rather certain that this hypothesis can be ruled out. Our analysis of the orbital poles therefore exacerbated the numerous problems which the $\Lambda$CDM model has in reproducing the observed MW satellite galaxy system \citep{Kroupa2010,Kroupa2012a,Kroupa2012b}. At the same time, our results reinforce the scenario in which the VPOS of satellites around the MW is populated by second-generation galaxies which were born as TDGs in a past galaxy encounter \citep{Pawlowski2011,Pawlowski2012a,Fouquet2012}. This realization finds further support in the discovery of the GPoA, a similar co-orbiting plane of satellite galaxies around the Andromeda galaxy \citep{Ibata2013,Conn2013,Hammer2013}. It is rather remarkable that the two only satellite galaxy systems for which three-dimensional positions are available both contain rotationally supported vast disc structures.

\section*{Acknowledgements}
M.S.P. acknowledges support through DFG reseach grant KR1635/18-2, in the frame of the \textit{DFG Priority Programme 1177, ``Witnesses of Cosmic History: Formation and evolution of galaxies, black holes, and their environment''}.
We thank Slawomir Piatek and Carlton Pryor for providing us with an updated preliminary PM for Draco.

\bibliographystyle{mn2e}
\bibliography{VPOSpropmo}

\label{lastpage}

\end{document}